\documentclass[fleqn,usenatbib]{mnras}

\usepackage{newtxtext,newtxmath}

\usepackage[T1]{fontenc}
\usepackage{ae,aecompl}

\usepackage{lineno}
\usepackage{gensymb}

\usepackage{graphicx}	
\usepackage{amsmath}	
\usepackage{amssymb}	

\newcommand\vx{\vec{x}}

\newcommand\Ng{N_{\rm g}}

\newcommand\oO{\mathcal{O}}

\newcommand\Vmax{V_{\rm max}}
\newcommand\Rmax{R_{\rm max}}

\newcommand\Mpc{\rm\; Mpc}
\newcommand\Mpch{{\rm\; Mpc}/h}

\newcommand{\Nbins}{N_{\rm bins}}

\title[Practical Computation of the Anisotropic 3PCF]{A Practical Computational Method for the Anisotropic Redshift-Space 3-Point Correlation Function}

\author[Slepian \& Eisenstein]{
Zachary Slepian$^{1,2}$\thanks{E-mail: zslepian@lbl.gov (ZS)} \&
Daniel J. Eisenstein$^{3}$\thanks{E-mail: deisenstein@cfa.harvard.edu (DJE)}
\\
$^{1}$Einstein Fellow, Lawrence Berkeley National Laboratory, 1 Cyclotron Road, Berkeley, CA 94720, USA\\
$^{2}${Berkeley Center for Cosmological Physics, University of California, Berkeley, CA 94720, USA}\\
$^{3}$Harvard-Smithsonian Center for Astrophysics, 60 Garden Street, Cambridge, MA 02138, USA
}

\date{Accepted XXX. Received YYY; in original form ZZZ}

\pubyear{2017}

\begin{document}
\label{firstpage}
\pagerange{\pageref{firstpage}--\pageref{lastpage}}
\maketitle

\begin{abstract}
We present an algorithm enabling computation of the anisotropic redshift-space galaxy 3-point correlation function (3PCF) scaling as $N^2$, with $N$ the number of galaxies. Our previous work showed how to compute the isotropic 3PCF with this scaling by expanding the radially-binned density field around each galaxy in the survey into spherical harmonics and combining these coefficients to form multipole moments. The $N^2$ scaling occurred because this approach never explicitly required the relative angle between a galaxy pair about the primary galaxy. Here we generalize this work, demonstrating that in the presence of azimuthally-symmetric anisotropy produced by redshift-space distortions (RSD) the 3PCF can be described by two triangle side lengths, two independent total angular momenta, and a spin. This basis for the anisotropic 3PCF allows its computation with negligible additional work over the isotropic 3PCF. We also present the covariance matrix of the anisotropic 3PCF measured in this basis. Our algorithm tracks the full 5-D redshift-space 3PCF, uses an accurate line of sight to each triplet, is exact in angle, and easily handles edge correction. It will enable use of the anisotropic large-scale 3PCF as a probe of RSD in current and upcoming large-scale redshift surveys.
\end{abstract}




\section{Introduction}
Measuring the clustering of galaxies is a standard cosmological probe, revealing the Universe's contents and laws while simultaneously illuminating the process of galaxy formation. This clustering is often quantified via correlation functions, which measure the excess above random of e.g. pairs (2-Point Correlation Function, 2PCF) or triplets (3PCF) of galaxies (\citealt{Peebles}). The Universe is homogeneous and isotropic on large scales, so in principle the correlation functions should be direction-independent (isotropic). However, in practice  spectroscopic surveys differ in their method for measuring objects' line-of-sight positions as opposed to the two transverse coordinates so the correlation functions become direction-dependent (anisotropic). 

In particular, an object's angular position is directly observable. In contrast, an object's line of sight position is inferred from its redshift by presuming that the redshift is due solely to the object's recession as it comoves with the background expansion of the Universe. But large-scale structure grows by convergence of matter onto overdense regions, generating peculiar velocities relative to the background expansion and rendering this assumption inaccurate. Furthermore, on smaller scales, virial motions of galaxies inside clusters generate additional peculiar velocities. The resulting patterns in the observed clustering, which on small scales ($\lesssim 20 \Mpc$) are parallel to the line of sight (``fingers of God''; \citealt{Jackson1972}) and on larger scales are transverse (``Kaiser pancakes''; \citealt{Kaiser1987}), are termed redshift-space distortions (RSD; \citealt{Hamilton1998}, for a review). 

Despite rendering cosmological parameter inference from galaxy clustering more difficult, RSD contain additional information on the laws of physics and the Universe's contents. In particular, they scale as $f=d\ln D/d\ln a\approx \Omega_{\rm m}^{\gamma}$, with $D$ the linear growth rate, $a$ the scale factor, $\Omega_{\rm m}$ the matter density observed at a particular redshift, and $\gamma = 0.56$ in General Relativity (GR) but with other values for alternative theories of gravity (\citealt{Linder05}). If GR is assumed, then RSD probe the matter density, and if the matter density is assumed, RSD test GR (e.g. \citealt{Raccanelli2013}).

RSD generate dependence of the observed clustering on angle with respect to the line of sight or lines of sight to a given $N$-tuplet of galaxies. In the simplest treatment of RSD, which uses linear perturbation theory and assumes a single line of sight to the entire survey (``flat sky''), RSD produce a quadrupole and hexadecapole $(\ell = 2$ and $\ell=4$) in this angle's cosine in the 2PCF or its Fourier analog the power spectrum (\citealt{Kaiser1987}; \citealt{Hamilton1993}).

The 2PCF and power spectrum multipoles have been used with considerable success to additionally constrain the parameters already probed by the isotropic ($\ell=0$) 2PCF as well as to measure $f$ (e.g. \citealt{Ross2017}; \citealt{Beutler2017}). Further, the multipoles can be integrated over and summed to produce clustering wedges in angle to the line of sight, offering an alternative route to parameter constraints (\citealt{Kazin2012}; \citealt{Grieb2017}; \citealt{Hand2017}). Models for RSD that go beyond linear perturbation theory have also been developed (\citealt{Taruya2010}; \citealt{Wang2014}; \citealt{Jeong2015}; \citealt{Bianchi2016}; \citealt{Vlah2016}; see \citealt{White2015} for comparison of recent models). Considerable work has also been done on wide-angle effects (e.g. \citealt{Raccanelli2016}; \citealt{Samushia2015}; \citealt{SE_aniso_wa}; \citealt{Papai2008}; \citealt{Reimberg2016}).

The situation for the 3-point correlation function (3PCF), measuring excess triplets of galaxies above random, is more complicated. The RSD now depend on the angles of two triangle sides to the line of sight (the angle of the third will be fixed by these two). Including the three parameters required to describe the triangle itself, there are now five free parameters even in the simplest flat-sky case (\citealt{Scoccimarro1999}). In principle, the anisotropic 3PCF contains rich additional information on galaxy biasing and the growth rate, but in practice, it is difficult to measure. Further, it is challenging to report and visualize, as it depends on five parameters.

The isotropic 3PCF already is computationally expensive, scaling as $N^3$ in the simplest approach, with $N$ the number of galaxies. To reduce this computational cost, different acceleration schemes have been proposed based on kd-trees and introducing approximations in how the clustering is measured (\citealt{Zhang2005}; \citealt{Gardner2007}; \citealt{March2013}). 

Our own recent work presented a 3PCF algorithm exploiting properties of the multipole basis that we believe is transformatively fast for large-scale 3PCF work, scaling as $N^2$ rather than $N^3$ \citep{SE_3pt_alg}. The algorithm further accelerates to become order $\Ng \log \Ng$, with $\Ng$ the number of grid points if Fourier Transforms are used \citep{SE_3pt_FT}. We showed that the multipole framework also allows accurate modeling of the covariance on large scales. We applied this framework to the Baryon Oscillation Spectroscopic Survey (BOSS) Data Release (DR) 12 Constant Mass (CMASS) sample of $\sim\!800,000$ Luminous Red Galaxies (LRGs) to detect the Baryon Acoustic Oscillations (BAO) (\citealt{SEcomp}, \citealt{SE_Full3PCF_BAO}; for pedagogical presentation of the BAO physics, see \citealt{SE_TF} and \citealt{ESW07}) as well as to constrain novel forms of biasing (\citealt{SE_Full3PCF_BAO}, \citealt{SE_RV_constraint}).  Additional work explored the theoretical predictions for the isotropically-averaged 3PCF in the multipole basis (\cite{RSD_model}) and provided the models fit in these observational works.

In this paper, we extend our 3PCF algorithm to track the line-of-sight dependence of the 3PCF. We show that a simple promotion of the Legendre coefficients relevant for the isotropic 3PCF to mixed spherical harmonic coefficients depending on two total angular momenta $l$ and $l'$ and one spin $m$ fully captures anisotropic clustering. These coefficients can be easily obtained using the same procedure previously developed for the isotropic 3PCF, which centered on obtaining spherical harmonic expansions of the density field on spherical shells around every galaxy in the survey. Further, adding a simple rotation of coordinates so that the position vector of the central galaxy serves as the line of sight to the triplet allows use of a varying line of sight, which tracks the anisotropic clustering more accurately than assuming a single line of sight to the entire survey.  Additionally, as shown in \cite{SE_3pt_FT}, the spherical harmonic coefficients can be obtained in $\Ng \log \Ng$ time using Fourier Transforms, and so if the density field is gridded the anisotropic 3PCF algorithm of this paper can be accelerated even further. 

In addition to the speed of measurement it enables, the basis advanced in this paper has two other important advantages. First, the parametrization of the 3PCF it involves, as coefficients depending on angular momenta $l, l'$, spin $m$ and triangle side lengths $r_1, r_2$, can be easily sliced for analysis and visualization. One might fix the angular momenta and spin and show a color plot versus $r_1$ and $r_2$, generalizing what was done in \cite{SE_3pt_alg}, \cite{SE_RV_sig}, and \citet{RSD_model} to look at the scale-dependent structure. Alternatively, one could examine the angular structure by fixing $r_1$,  $r_2$, and $l$ and showing the dependence on $l'$ and $m$.  

Second, our parametrization permits straightforward handling of the covariance matrix of the anisotropic 3PCF. We extend the work of \cite{SE_3pt_alg} to compute the anisotropic
covariance matrix assuming a boundary-free survey whose density
perturbations follow a Gaussian Random Field (GRF) with an
anisotropic power spectrum given by the Kaiser formula for RSD. We leave the survey volume and the number density as free parameters to be fit from mock catalogs. This calculation supplies an important advance: having a smooth covariance matrix means it can be inverted, a known difficulty for covariance matrices estimated from large numbers of mock catalogs (\citealt{Percival2014}). In particular, the inverse of the covariance depends on the smallest eigenvalue and to accurately obtain this one requires many mock catalogs per dimension of the covariance matrix. Since the dimension can be large, one often requires the 3PCF for thousands of mocks, adding significant time to any analysis.  

In contrast, given our template with only two free parameters, one does not require many mocks to fit for them and obtain a smoothly invertible covariance. The basis we propose permits straightforward computation of this template covariance. In particular, expanding in angular momentum eigenstates (which the spherical harmonics are) means that the angular integrals to bring the covariance from Fourier space (where the GRF calculation is easiest) to configuration space (where the measurement is done) simplify greatly.  These integrals, formally 12-D, can be reduced to 1-D and 2-D integrals of the power spectrum, enabling fast evaluation on a grid rather than using more complicated higher-dimensional integration techniques.

The paper is laid out as follows. \S\ref{sec:basis} presents the basis we will use for our algorithm and shows how it emerges from imposing symmetry about the line of sight (taken to be the $z$ axis) on the most general representation for the two vectors defining a given triangle. \S\ref{sec:algorithm} outlines the algorithm, showing that only a slight generalization of \cite{SE_3pt_alg} is needed to obtain the anisotropic 3PCF and that the speed remains $\oO(N^2)$. \S\ref{sec:los} describes how a varying line of sight that follows each galaxy triplet can be incorporated. \S\ref{sec:edge} discusses edge correction in the algorithm's basis. In \S{\ref{sec:covar} we compute the covariance of the anisotropic 3PCF in the limited but useful approximation described above. \S\ref{sec:conclusions} concludes and is followed by two brief Appendices, one with identities used in the work's main body (\nameref{sec:appendix_A}) and a second showing the impossibility of using a triple Legendre series for the anisotropic 3PCF (\nameref{sec:appendix_B}).

\section{Basis}
\label{sec:basis}
Consider a triplet of galaxies at positions $\vec{x}$, $\vec{x}+\vec{r}_1$, and $\vec{x}+\vec{r}_2$. An estimate of the full 3PCF about $\vx$ including any possible dependence on the triangle configuration as well as on its orientation is 
\begin{align}
&\hat{\zeta}(\vec{r}_1,\vec{r}_2;\vec{x}) = \nonumber\\
&\frac{4\pi}{\sqrt{(2l+1)(2l'+1)}}\sum_{lm} \sum_{l'm'} \hat{\zeta}_{l l'}^{m m'} (r_1, r_2; \vec{x})Y_{lm}(\hat{r}_1) Y^*_{l'm'}(\hat{r}_2).
\label{eqn:full_3pcf_est}
\end{align}
The $Y_{lm}$ are spherical harmonics and the pre-factor is a normalization to recover the 3PCF's expansion into Legendre polynomials in the isotropic limit as in \cite{SE_3pt_alg}.

Since RSD are due to the difference between how the line of sight and transverse positions are computed from a survey, they must be symmetric under rotations about the line of sight $\hat{n}$. Here we take $\hat{n}= \hat{x}$, i.e. that there is a single line of sight to the entire triangle, given by the line of sight to the galaxy at the vertex where the triangle's opening angle is defined. This geometry is shown in Figure \ref{fig:basis_approach}, and we further discuss this choice for the line of sight in \S\ref{sec:los}. We work in a coordinate system where the $z$ axis is along $\hat{x}$. Averaging over rotations around $\hat{n} = \hat{x} = \hat{z}$, the azimuthally-averaged 3PCF, denoted with subscript ``azi'', is
\begin{align}
&\hat{\zeta}_{\rm azi}(\vec{r}_1,\vec{r}_2;\vec{x}) = \int_{0}^{2\pi} d\phi \;\hat{\zeta}(\vec{r}_1,\vec{r}_2;\vec{x})\nonumber\\
&=\frac{4\pi}{\sqrt{(2l+1)(2l'+1)}}\sum_{lm} \sum_{l'm'} \hat{\zeta}_{l l'}^{m m'} (r_1, r_2; \vec{x})\nonumber\\
&\times \int_{0}^{2\pi} d \phi \;Y_{lm} ({\bf R}_{z}(\phi) \hat{r}_1) Y^*_{l'm'} ({\bf R}_{z}(\phi)\hat{r}_2),
\label{eqn:azi_setup}
\end{align}
where ${\bf R}_{z}(\phi)$ represents a rotation by an angle $\phi$ about $\hat{z}$. The spherical harmonics are defined 
\begin{equation}
Y_{lm}(\theta,\phi)=\sqrt{\frac{2l+1}{4\pi}\frac{(l-m)!}{(l+m)!}}P_{l}^{m}(\cos\theta)e^{im\phi},
\label{eqn:ylm_defn}
\end{equation}
so the only $\phi$ dependence is in the exponential.

We can set the initial azimuthal angle $\phi_1$ of $\hat{r}_1$ to zero, and the initial angle of $\hat{r}_2$ we denote $\phi_2$. Applying the rotation ${\bf R}_{z}(\phi)$ then simply adds an angle $\phi$ to each azimuthal angle, so the azimuthal average of the spherical harmonics in equation (\ref{eqn:azi_setup}) scales as
\begin{align}
&\int_{0}^{2\pi} d \phi \;Y_{lm} ({\bf R}_{z}(\phi) \hat{r}_1) Y^*_{l'm'} ({\bf R}_{z}(\phi)\hat{r}_2) \propto \int_{0}^{2\pi} d\phi\;  e^{i m \phi} e^{-i m' (\phi_2 + \phi)}\nonumber\\
&\propto \int_{0}^{2\pi} d\phi\;  e^{i (m-m')\phi} = \delta^K_{m m'},
\end{align}
where in the second line we took the exponential in $\phi_2$ outside the integral and then dropped it. $\delta^K_{m m'}$ is a Kronecker delta, unity if the subscripted arguments are equal and zero otherwise. 

Inserting this result in equation (\ref{eqn:full_3pcf_est}), we see that only spherical harmonic combinations where $m=m'$ can enter the azimuthally-averaged 3PCF.  We note that $l$ need not equal $l'$, in contrast to the isotropic case where averaging the spherical harmonics over full 3-D rotations forces $l= l'$ as well (\citealt{SE_3pt_alg}). Thus, the azimuthally-averaged 3PCF estimate is fully described as
\begin{align}
&\hat{\zeta}_{\rm azi}(\vec{r}_1,\vec{r}_2;\vec{x}) =\nonumber\\
&\frac{4\pi}{\sqrt{(2l+1)(2l'+1)}} \sum_{l l'} \sum_{m} \hat{\zeta}_{l l'}^{m} (r_1, r_2; \vec{x})Y_{lm}(\hat{r}_1) Y^*_{l' m}(\hat{r}_2);
\label{eqn:azi_3pcf_est}
\end{align}
moving forward we discuss only the azimuthally-averaged 3PCF and so we suppress the subscript ``azi.'' Integrating over $d^3\vec{x}$ yields the full 3PCF $\zeta$ as the translation average of the estimate about a particular galaxy at $\vec{x}$ given by equation (\ref{eqn:azi_3pcf_est}).

We now discuss an additional symmetry relevant for this basis. We need only compute the coefficients $\zeta_{ll'}^m$ for $m\geq 0$ as those for $m<0$ are the complex conjugate of the $m>0$ coefficients. There are several ways to understand this point.

First, mathematically, we have
\begin{align}
\zeta_{ll'}^m(r_1, r_2;\vec{x})=\int d\Omega_1 d\Omega_1\;\hat{\zeta}(r_1, r_2)Y_{lm}^*(\hat{r}_1) Y_{l' m}(\hat{r}_2)
\label{eqn:zeta_plus_m}
\end{align}
and
\begin{align}
\zeta_{ll'}^{-m}(r_1, r_2;\vec{x})=\int d\Omega_1 d\Omega_1\;\hat{\zeta}(r_1, r_2)Y_{l-m}^*(\hat{r}_1) Y_{l' -m}(\hat{r}_2)
\label{eqn:zeta_minus_m}
\end{align}
Using the identity that $Y^*_{lm}(\hat{r}) = (-1)^m Y_{l-m}(\hat{r})$, equation (\ref{eqn:zeta_minus_m}) becomes
\begin{align}
\zeta_{ll'}^{-m}(r_1, r_2;\vec{x}) &= (-1)^{2m}\int d\Omega_1 d\Omega_1\;\hat{\zeta}(r_1, r_2)Y_{lm}(\hat{r}_1) Y^*_{l' m}(\hat{r}_2)\nonumber\\
&=\left(\zeta_{ll'}^m(r_1, r_2;\vec{x}) \right)^*
\end{align}
where the second equality follows by noting that $(-1)^{2m} = 1$, conjugating equation (\ref{eqn:zeta_plus_m}), and recalling that $\zeta(\vec{r}_1, \vec{r}_2; \vec{x})$ is real.

Physically, the redundancy of the $-m$ coefficients occurs because flipping $m \to -m$ is equivalent to flipping the azimuthal angle $\phi \to -\phi$, modulo a factor of $(-1)^m$.\footnote{Flipping $m$ also affects the associated Legendre polynomial in equation (\ref{eqn:ylm_defn}) because $P_l^{-m}(\cos \theta) = (-1)^m\left[(l-m)!/(l+m)!\right] P_l^m (\cos \theta)$, but the factorial piece here cancels the factorials in the spherical harmonic's definition after $m\to -m$ is taken there. Consequently flipping $m$ in the associated Legendre polynomial only contributes an overall factor of $(-1)^m$ to the flipped-$m$ spherical harmonic.} Considering $\phi$ to be defined as the angle swept out as one moves from the $x$-axis towards the $y$-axis in the $xy$-plane, this transformation is equivalent to flipping $y\to -y$ while keeping the $x$ and $z$ axes fixed.\footnote{This interpretation is easily manifested by writing the spherical harmonics in the Cartesian basis, where they are proportional to powers of $(x+iy)/r$ for $m>0$ and $(x-iy)/r$ for $m<0$; this of course just comes from applying Euler's formula to $\exp[i m\phi]$ and identifying $\cos \phi = x/r$ and $\sin \phi =y/r$.} This flip corresponds to reversing the handedness of the coordinate system, as it now satisfies a left-hand rule ($\hat{x}\times \hat{y} = -\hat{z}$) rather than a right-hand rule. We see then that the redundancy of the $-m$ coefficients reflects the physical symmetry that the galaxy distribution about any point $\vec{x}$ is insensitive to the handedness of the coordinate system one chooses around that point.

As a result of this symmetry, in the sum over $m$ of equation (\ref{eqn:azi_3pcf_est}) the negative and positive $m$ can always be paired to give a real result that is $2\;{\rm Re} \zeta_{ll'}^m$. Consequently, in what follows we will have it in mind that the final results reported from our approach to the anisotropic 3PCF will be real and symmetrized (denoted by a bar) over positive and negative spins as
\begin{align}
\bar{\zeta}_{ll'}^m\equiv \zeta_{ll'}^m + \zeta_{ll'}^{-m}(1-\delta^{\rm K}_{m0}) &= (2 - \delta^{\rm K}_{m0})\;{\rm Re}\;\zeta_{ll'}^m
\label{eqn:zeta_symm_defn}
\end{align} 
where we take $m \geq 0$ above. However, throughout the paper we will often find it convenient to perform analytic calculations in terms of the unsymmetrized $\zeta_{ll'}^m$ and symmetrize at the final step.

We now discuss the behavior of $\zeta_{ll'}$ under parity. $\zeta_{ll'}^m$ behaves as $(-1)^{l+l'}$ under this transformation. For indistinguishable points, such as galaxies from a single population or survey, the anisotropic 3PCF must be symmetric under parity, meaning $l+l'$ must be even. $l+l'$ need not be even for e.g. a 3PCF formed from taking two points from one galaxy population and a third from a different population at higher redshift, as this choice would introduce a preferred orientation.

Finally, in the isotropic limit, $l'=l$ and the coefficient $\zeta_{ll'}^m$ becomes $m$-independent, i.e. $\zeta_{ll'}^m\to \zeta_l$. We can then sum the spherical harmonics in equation (\ref{eqn:azi_3pcf_est}) over spins using the spherical harmonic addition theorem (\citealt{AWH13} equation 16.57) to recover that $\zeta(r_1, r_2;\hat{r}_1\cdot\hat{r}_2;\vec{x}) = \sum_l \zeta_l(r_1, r_2;\vec{x})\mathcal{L}_l(\hat{r}_1\cdot\hat{r}_2)$ as in \cite{SE_3pt_alg}, with $\mathcal{L}_l$ a Legendre polynomial of order $l$.

\begin{figure}
\includegraphics[scale=0.35]{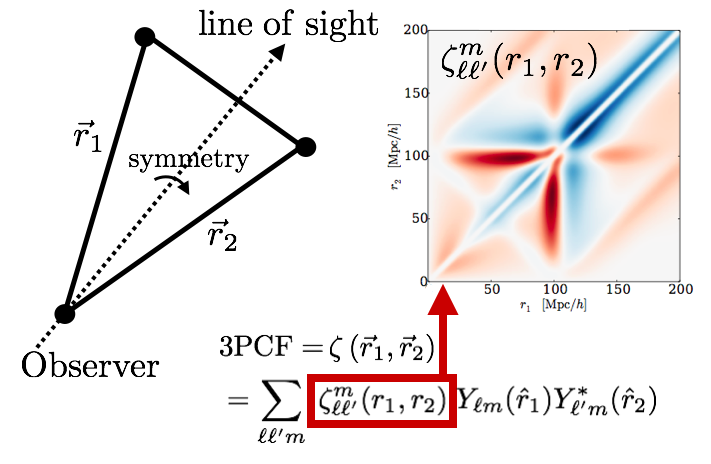}
\caption{Here we schematically show the approach of this paper. The 3PCF is parametrized by two vectors $\vec{r}_1$ and $\vec{r}_2$, which encode the triangle's side lengths and its orientation with respect to the line of sight, shown in dashed black. The azimuthal symmetry about the line of sight is the reason only a single spin $m$ enters the basis. The basis is shown in the red-boxed equation: we expand the anisotropic 3PCF in radial coefficients $\zeta_{ll'}^m$ times spherical harmonics in $\hat{r}_1$ and $\hat{r}_2$. We seek to measure the radial coefficients and can then plot them at fixed $l, l'$, and $m$ versus $r_1$ and $r_2$ to reveal the spatial structure. Alternatively, we might also plot the coefficients at fixed $r_1, r_2$, and $l$ versus $l'$ and $m$ to show the angular structure. The features are BAO, and the plot is adopted from \citet{RSD_model} just to suggest how these plots might look.}
\label{fig:basis_approach}
\end{figure}

\section{Algorithm}
\label{sec:algorithm}
We now show how the anisotropic 3PCF coefficients in the spherical harmonic basis may be obtained in $\mathcal{O}(N^2)$ time with $N$ the number of galaxies.  Here we focus on the 3PCF of an arbitrary density field given by $\delta$; in \S\ref{sec:edge} we will discuss applying this approach to the anisotropic analog of the \citet{Szapudi1998} estimator for the isotropic 3PCF. 

We again begin with a ``primary'' galaxy at $\vec{x}$ and denote the positions of two ``secondary'' galaxies as $\vec{x}+\vec{r}_1$ and $\vec{x}+\vec{r}_2$. This configuration forms a triangle; we bin its side lengths into radial bins denoted $r_1$ and $r_2$. We denote the radially binned density field 
\begin{align}
\bar{\delta}(r_i; \hat{r}_i;\vec{x}) = \int r^2 dr\; \Phi(r;r_i) \delta(\vec{x}+ \vec{r}),
\end{align}
where $\Phi$ is a binning function that ensures $r$ is in the bin denoted by $r_i$.

We now desire an estimate of the anisotropic binned 3PCF coefficients about $\vec{x}$.  We have
\begin{align}
&\hat{\zeta}_{l l'}^{m}(r_1, r_2;\vec{x}) = \nonumber\\
&\delta(\vec{x}) \int d\Omega_1 d\Omega_2\; \bar{\delta}(r_1; \hat{r}_1;\vec{x}) \bar{\delta}(r_2; \hat{r}_2;\vec{x}) Y_{lm}(\hat{r}_1)Y_{l'm}^*(\hat{r}_2)
\end{align}
It is immediate that this double integral factorizes; defining spherical harmonic coefficients of the binned density as
\begin{align}
a_{lm}(r_i; \vec{x}) = \int d\Omega \;Y_{lm}^*(\hat{r}) \bar{\delta}(r_i; \hat{r}_i; \vec{x})
\end{align}
we may write
\begin{align}
\hat{\zeta}_{l l'}^{m}(r_1, r_2;\vec{x}) &= \delta(\vec{x}) a_{lm}(r_1;\vec{x})a^*_{l'm}(r_2;\vec{x}).
\label{eqn:zeta_ito_alm}
\end{align}
Symmetrizing and using that $\hat{\zeta}_{ll'}^{-m} = \left(\hat{\zeta}_{ll'}^m\right)^*$, we find that
\begin{align}
&\hat{\bar{\zeta}}_{ll'}^m(r_1, r_2;\vec{x}) \nonumber\\
&= \delta(\vec{x})\big[a_{lm}(r_1;\vec{x})a_{l'm}^*(r_2;\vec{x}) + a_{lm}^*(r_1;\vec{x})a_{l'm}(r_2;\vec{x})\big],
\label{eqn:zeta_bar_ito_alm}
\end{align}
which is manifestly symmetric under flipping the conjugate signs, and so our choice to place the conjugate on $a_{lm}(r_1)$ in equation (\ref{eqn:zeta_ito_alm}) for the $\hat{\zeta}_{ll'}^m$ used to construct $\hat{\bar{\zeta}}_{ll'}^m$ did not matter.

Importantly, computing the $a_{lm}$ about a given primary galaxy at $\vec{x}$ on a given radial bin scales as the number of galaxies in that bin. In total if we compute correlations out to a radius $R_{\rm max}$, obtaining the $a_{lm}$ on all bins scales as $nV_{\rm max}$ with $n$ the survey number density and $V_{\rm max}$ the volume of a sphere with radius $R_{\rm max}$.  The $a_{lm}$ must be obtained around every galaxy, so the total work scales as $N\left(nV_{\rm max}\right)$. 

In detail, around each primary,  at each $l$ we have $l + 1$ distinct $a_{lm}$, as we require only the $m \geq 0$ spherical harmonic coefficients due to the symmetry that $\hat{\zeta}_{ll'}^{-m} = \left(\hat{\zeta}_{ll'}^m\right)^*$ as discussed in \S\ref{sec:basis}. The total number of coefficients up to $l_{\rm max}$ is then $(l_{\rm max}+2)(l_{\rm max}+1)/2$. Each of these must be computed on each bin, so the total number of coefficients to store is $\Nbins (l_{\rm max}+2)(l_{\rm max}+1)/2$. We note that the total work of obtaining these scales as $(nV_{\rm max})(l_{\rm max}+2)(l_{\rm max}+1)/2$, as the first factor accounts for all of the bins out to $R_{\rm max}$. 

We now briefly compute the number of combinations of these coefficients that must be formed around each primary. While forming these combinations is a negligible fraction of the total work (we find $\sim 2\%$ for the implementation discussed in \S\ref{sec:conclusions}), for completeness we discuss the combinatoric computation as it involves several steps. We  consider $l\geq l'$ without loss of generality and focus on $l>l'$ first. At each $l$, we have $l$ allowed values of $l'$, and at each $l'$, we have $l'+1$ values of $m$. We must now sum this over $l'$ up to $l$ and then $l$ from $0$ to $l_{\rm max}$. We find 
\begin{align}
N_{{\rm combs,\;} l>l'}= \Nbins^2\sum_{l=0}^{l_{\rm max}}\sum_{l'=0}^{l-1} (l'+1) = \frac{\Nbins^2}{2}l_{\rm max}(l_{\rm max}+1)^2,
\label{eqn:not_equal}
\end{align}
noting that we include $\Nbins^2$ rather than $(\Nbins+1)\Nbins/2$ because there is no switch symmetry of the bins as $l\neq l'$.

For the $l=l'$ piece, we have just one allowed $l'$ at each $l$, but still $l+1$ allowed spins $m$, but there is now a switch symmetry between the bins,  so we compute 
\begin{align}
N_{{\rm combs}, l=l'} &= \frac{(\Nbins+1)\Nbins}{2}\sum_{l=0}^{l_{\rm max}}(l+1)\nonumber\\ &=\frac{(\Nbins+1)\Nbins}{4}(l_{\rm max}+1)^2.
\label{eqn:equal}
\end{align}
The full number of harmonic coefficient combinations required is then the sum of equations (\ref{eqn:not_equal}) and (\ref{eqn:equal}).
\section{Determination of the Line of sight}
\label{sec:los}
\subsection{A generalized Yamamoto estimator}
In this work we take the line of sight to a given triangle of galaxies to be the position vector $\vec{x}$ of the galaxy at which the triangle's opening angle is defined.  This galaxy is the ``primary,'' and serves both to define the origin of coordinates and the $z$ axis for computing the spherical harmonic expansion of the radially-binnned ``secondary'' galaxies around it.

Defining the line of sight as the position vector of one triplet member is the natural generalization of the \cite{Yamamoto2006} estimator for the anisotropic 2PCF or power spectrum (further discussed in this latter context in \citealt{SE_3pt_FT}, \citealt{Bianchi_cart_2015}, and \citealt{Scoccimarro2015}). As shown in \cite{SE_aniso_wa} for the anisotropic 2PCF, wide angle effects enter this estimator only at $\oO(\theta^2)$, where $\theta$ is the opening angle of the triangle formed by the observer and the galaxy pair. 

Further, the Yamamoto estimator differs from using the angle bisector of this triangle or the midpoint of the pair separation only at $\oO(\theta^2)$.  The cancellation of the $\oO(\theta)$ effect occurs because it has odd parity. Thus when the effect is summed over the two options for the line of the sight to a galaxy pair (the position vectors of the first and the second pair members), it cancels \citep{SE_aniso_wa}.

These arguments rely on expanding the anisotropic 2PCF in Legendre polynomials tracking the cosine of the angle between the pair separation and the line of sight. We suspect that in the current work cancellation of the $\oO(\theta)$ error will also occur. In any case, using a rotating line of sight for the anisotropic 3PCF is undoubtedly more accurate than the flat-sky approximation. Finally, for completeness we note that there has been one other work using a rotating line of sight for 3-point clustering: \cite{Scoccimarro2015} presents a rotating line of sight estimator for multipole moments of the bispectrum averaged over rotations about one of the wave-vectors.  

\begin{figure}
\includegraphics[scale=0.208]{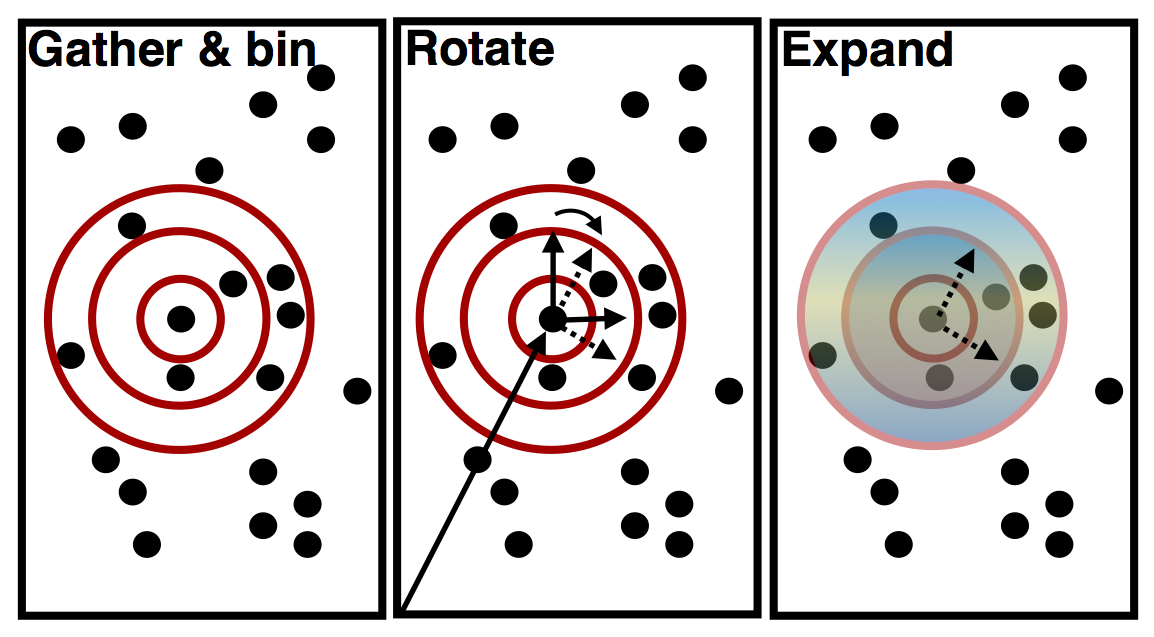}
\caption{Here we show the algorithm flow around a given primary at a point $\vec{x}$ for the pre-rotation approach. First, we gather all galaxies within $\Rmax$ and bin them into spherical shells (also called radial bins). Then we rotate so that the primary lies along the $z$-axis. Finally, on each spherical shell we expand the angular dependence of the density into spherical harmonics with coefficients $a_{lm}(r_i;\vec{x})$, with $r_i$ designating the radial bin.}
\label{fig:alg_flow}
\end{figure}

\subsection{Computing using the rotating line of sight}
\label{subsec:rotating}
There are two options for computing the 3PCF using the moving line of sight outlined above. Which option is preferable is likely use-case dependent, so we present both. 

The simpler approach is to rotate the coordinates of all the secondaries about a given primary into the frame where the primary lies on the $z$ axis. We term this ``pre-rotation.'' The computational cost here about a given primary is the number of secondaries, $n\Vmax$, and the rotation must be performed around each of the $N$ primaries, so the total cost is $N(n\Vmax)=\oO(N^2)$, the same scaling for this approach as for the overall 3PCF algorithm. The algorithm flow here is shown in Figure \ref{fig:alg_flow}. We note that while the scaling is formally $\oO(N^2)$, the rotations are still a  small amount of work relative to the multipole computation, as we further detail below. 

We now outline how the pre-rotation computation proceeds. We rotate the primary and all secondaries to a system where the primary is along the $z$-axis, which is the line of sight. We define a triad of unit vectors, with one being the line of sight and the other two orthogonal to it, and then 
dot them into the separation vectors between the secondaries and the primary. This yields the secondaries' new coordinates in our desired system. This procedure requires just one matrix by vector multiply (9 multiply-adds), as compared to the 286 multiply-adds required to then obtain the multipole contributions for a given secondary (with $\ell_{\rm max} = 10$). Thus pre-rotation only adds $3$\% more work for a given secondary. Further, it is independent of the number of multipoles and number of radial bins.

We now discuss a second approach to rotation. We observe that rotations will not mix radial bins; a secondary in a given radial bin will be in the same bin post-rotation.  Further, $l_{\rm max}$ sets the fineness with which the angular structure of the secondaries on a given spherical shell is probed; much as in the Cosmic Microwave Background (CMB) an angular momentum $l$ probes angular separations as $\Delta \theta \sim\!180^{\circ}/l$.  Rotations do not change the underlying angular structure of the secondaries on a shell, meaning that they should not mix information from different $l$, and in particular that measuring a given $l_{\rm max}$ in any basis offers the same $l_{\rm max}$ after rotation. However, the information in the spins $m$ will mix. Rotating about the $z$ axis means a given frequency $m$ (recall $Y_{lm}\propto \exp[i m\phi]$) in the original system will map to a new frequency $m'$ in the new, primed system. This mapping occurs because the new plane in which $\phi'$ is measured will be at some angle to the original plane in which $\phi$ was measured, and therefore a sinusoid with frequency $m$ in the original plane will have a new frequency when it is measured in projection. Indeed, in the limiting case of a rotation by $90^{\degree}$, a displacement by $\Delta \phi$ in the old system would map to a displacement by $\Delta \phi' = 0$ in the new system, entirely changing the spin structure of the expansion.

These points suggest that we can reconstruct the spherical harmonic coefficients in our rotated basis from a sum over spins of those measured in any other basis, in particular whatever ``global'' basis in which all $x, y,$ and $z$ galaxy positions in a survey might be specified.  In this section only we adopt the notation superscript ``G'' to denote the spherical harmonic coefficients measured in this ``global'' basis, and superscript ``L'' to denote those in the local basis rotated about a primary at $\vec{x}$ so that $\vec{x}$ is along the $z$-axis.

To derive the relation between the ``global'' and ``local'' coefficients, consider the density field $\delta$ on a shell denoted by $r$ about a primary at $\vec{x}$:
\begin{align}
\delta(r;\hat{r};\vec{x}) = \sum_{LM} a^G_{LM}(r;\vec{x})Y_{LM}(\hat{r}).
 \end{align} 
Under the desired rotation ${\bf R}$, the density becomes
\begin{align}
&\delta(r;{\bf R}\hat{r};\vec{x}) = \sum_{LM} a^G_{LM}(r;\vec{x})Y_{LM}({\bf R}\hat{r})\nonumber\\
&=\sum_{LM} a^G_{LM}(r;\vec{x}) \sum_{M'} D^L_{MM'}(\vec{x}) Y_{LM'}(\hat{r})
\label{eqn:global_exp}
\end{align} 
where in the second line we expanded the rotated spherical harmonic as a sum over unrotated spherical harmonics; $D^L_{MM'}$ is a Wigner D-matrix (e.g. \citealt{AWH13} equation (16.52) or \citealt{VMK06} chapter 4).

Noting that the expansion of the rotated density field in the ``local'' basis where we desire the coefficients is
\begin{align}
\delta(r;{\bf R}\hat{r};\vec{x}) = \sum_{lm} a^L_{lm}(r;\vec{x})Y_{lm}(\hat{r}),
\label{eqn:local_exp}
 \end{align} 
setting the expansions (\ref{eqn:global_exp}) and (\ref{eqn:local_exp}) equal, and invoking orthogonality, we find
\begin{align}
a^{L}_{lm}(r;\vec{x}) = \sum_M a^G_{l M}(r;\vec{x}) D^l_{Mm}(\vec{x}).
\label{eqn:glob_loc_reln}
\end{align}

With the relation (\ref{eqn:glob_loc_reln}), we can measure the spherical harmonic coefficients in any desired global basis and after finding them around all primaries, recombine locally on each bin about each primary weighted by the Wigner D-matrices. Since the rotations occur after the spherical harmonic coefficients are computed, we term this approach ``post-rotation.''

The scaling of this post-rotation approach differs from that of the pre-rotation approach outlined earlier, as we now show. We need only obtain $m\geq 0$ spherical harmonic coefficients as noted in \S\ref{sec:algorithm}, and for the same reason on the right-hand side we need only $a_{lM}$ with $M \geq 0$. Further, the D-matrices for $M<0$ can be related to those for $M>0$, and using symmetry properties of the D-matrix we can also ensure that $m\leq M$. Thus the entire right-hand side of equation (\ref{eqn:glob_loc_reln}) can be cast in terms of $M\geq 0$ spherical harmonic coefficients and D-matrices and using only D-matrices with $m\leq M$.  At each $l$ we then have $(l+1)(l+2)/2$ D-matrix elements to compute, leading to $286$ matrix elements for $l_{\rm max} = 10$.  The algorithm flow for this ``post-rotation'' approach is shown in Figure \ref{fig:alg_flow_alm_rot}.

A cross-check on this result is that all of the spherical harmonics up to $l_{\rm max}=10$ can be computed using 286 combinations of powers of $x/r,y/r$, and $z/r$ where $x,y,$ and $z$ are the relative coordinates of a secondary galaxy in the frame where the primary is at the origin (see \citealt{SE_3pt_alg}).  The behavior of these 286 fundamental power combinations under rotation must completely determine the rotated spherical harmonic expansion, confirming that we should need 286 matrix elements to perform the rotation.

The computational cost of evaluating the Wigner D-matrices is small for the modest $l$ we require. Direct evaluation is one option; for instance \cite{VMK06} gives easily-implemented expressions in terms of  Gauss's hypergeometric function $_2 F_1$. There also exist more efficient methods for their computation, such as the use of recursion relations or pseudo-spectral projection (primarily important going to high $l$) (\citealt{VMK06}; \citealt{Gimbutas2009}, \citealt{Gumerov2014}; \citealt{Feng2015}). 

We have written $D^l_{Mm}(\vec{x})$ as a function of the primary location $\vec{x}$. In detail, following the conventions of \cite{VMK06}, the D-matrix as a function of the Euler angles $\alpha = 0, \beta = -\theta(\vec{x}),$ and $\gamma=-\phi(\vec{x})$ is
\begin{align}
D^J_{MM'}(\alpha, \beta, \gamma) = e^{-i M\alpha} d_{MM'}^J(\beta) e^{-i M' \gamma},
\end{align}
where $d_{MM'}^J$ is a little-d matrix (see \citealt{VMK06}).

In contrast to pre-rotation, for post-rotation the total computational cost is independent of the number of secondaries about a given primary, but depends on the number of multipoles and number of radial bins, scaling as $l_{\rm max}^3\Nbins$.
\begin{figure}
\includegraphics[scale=0.28]{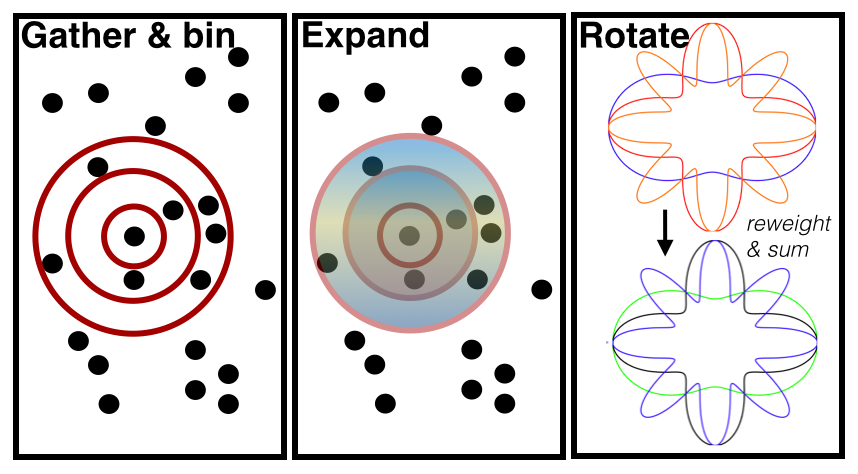}
\caption{Here we show the algorithm flow when rotation about a given primary is done after the spherical harmonic expansion rather than prior to it, and hence in the space of spherical harmonic coefficients $a_{lm}$ rather than spatial coordinates. We term this ``post-rotation.'' The ``Gather \& bin'' step is unchanged from Figure \ref{fig:alg_flow}, but we now expand the density's angular dependence into spherical harmonics on each bin in the second step rather than the third. We then take the $a_{lm}$ coefficients on a given bin (three are schematically indicated here in the different colors, with color denoting the coefficient's amplitude), reweight them as in equation (\ref{eqn:glob_loc_reln}) (new colors mean new amplitudes), and sum to obtain the new, rotated $a_{lm}$.}
\label{fig:alg_flow_alm_rot}
\end{figure}

\subsection{Using Fourier Transforms for the harmonic coefficients}
As shown in \cite{SE_3pt_FT}, the spherical harmonic coefficients about a given primary galaxy are simply a convolution integral, and so Fast Fourier Transforms (FFTs) can be used to obtain the coefficients about all primaries in the survey at once scaling as $\Ng \log \Ng$, with $\Ng$ the number of grid points, if the galaxy density field is gridded. While the fineness of grid required will be somewhat application-dependent, there are likely contexts (such as analysis of a large number of mocks for verifying the pipeline or covariance matrix) where a possible loss in precision will be offset by the increase in speed FFTs afford.  

Importantly, the post-rotation approach outlined in \S\ref{subsec:rotating} is essential for enabling use of FTs to obtain the spherical harmonics. The pre-rotation approach relies on sitting on a given primary galaxy and rotating using its coordinates, which is not possible in an FFT-based approach. However, once the spherical harmonic coefficients are known about each primary galaxy for any given choice of coordinates, as shown in \S\ref{subsec:rotating}, they can be rotated on a primary-by-primary basis with little additional computation. Thus, the coefficients can be obtained in some arbitrary global frame by FTs and then post-processed to give the desired identification of the local $z$-axis with the line of sight to each primary galaxy.

\section{Edge correction}
\label{sec:edge}
We now discuss edge correction in the spherical harmonic basis.  The estimator for the full 3PCF is \citep{Szapudi1998}
\begin{align}
\hat{\zeta}=\frac{NNN}{RRR},
\label{eqn:SS_est}
\end{align}
where $N\equiv D-R$ is the data minus random count with $D$ indicating data and $R$ indicating random. The interpretation of this estimator as an optimal weighting in the shot-noise limit and how it should be used to obtain the 3PCF is further discussed in \cite{SE_3pt_alg}.

\subsection{Edge correction in the spherical harmonic basis}
Here we show how to implement edge correction using our algorithm. We will begin with all spins, including negative ones, and then show how to cast the edge correction in terms of solely positive semi-definite spins so that we may work with the symmetrized 3PCF coefficients $\hat{\bar{\zeta}}){ll'}^m$. We multiply each side of the estimator (\ref{eqn:SS_est}) by $RRR$ and  then expand the result into spherical harmonics, obtaining
\begin{align}
&\sum_{jj' s} \mathcal{N}^s_{j j'} Y_{js}(\hat{r}_1) Y_{j's}^*(\hat{r}_2)\nonumber\\
& = \sum_{l l' m} \sum_{k k' p} \hat{\zeta}_{l l' }^m \mathcal{R}_{k k'}^p Y_{lm}(\hat{r}_1) Y_{kp}(\hat{r}_1) Y_{l' m}^*(\hat{r}_2) Y_{k' p}^*(\hat{r}_2),
\end{align}
where the $\mathcal{N}_{jj'}^s$ are the double spherical harmonic coefficients of $N$, $\mathcal{R}_{kk'}^{p}$ those for $R$, and $\hat{\zeta}_{ll'}^m$ those for $\hat{\zeta}$. Integrating both sides against $Y^*_{js}(\hat{r}_1) Y_{j' s}(\hat{r}_2)$ and invoking orthogonality we find
\begin{align}
\mathcal{N}_{jj'}^s = \sum_{l l' m} \sum_{k k' p} \hat{\zeta}_{l l'}^m \mathcal{R}_{k k'}^p \mathcal{G}_{l k j}^{m p -s} \mathcal{G}_{l' k' j'}^{-m -p s}
\label{eqn:edge_orthog}
\end{align}
where $\mathcal{G}$ is the Gaunt integral, i.e. the integral of three spherical harmonics with the indicated total angular momenta (subscripts) and spins (superscripts); we write it in terms of 3j-symbols in \nameref{sec:appendix_A}. We note that if $k=0=k'$, then $l=j$ and $l'=j'$: if the randoms have no angular structure, then the coefficients $\mathcal{N}_{jj'}^s$ are the desired coefficients $\hat{\zeta}$ of the 3PCF up to normalization. In this case the edge correction was simply a division by the average background count to convert the galaxy density field into an overdensity field. We further observe that $k+k'$ and $j+j'$ must have the same parity, because the Gaunt integrals enforce that $l+k+j$ and $l'+k'+j'$ are even, and $l+l'$ is even, so $k+k'+j+j'$ is even as well. This rule somewhat reduces the number of random coefficients $\mathcal{R}$ required to obtain a given 3PCF coefficient from the measured difference field $\mathcal{N} = N-R$. Unlike the 3PCF coefficients, the random coefficients need not be symmetric under parity because the survey geometry may not be. Hence, the difference field coefficients need not be either.

We now show how equation (\ref{eqn:edge_orthog}) can be recast in terms of the symmetrized quantities our algorithm tracks. We first rewrite each coefficient separated into its real and imaginary parts, giving
\begin{align}
\mathcal{N}_{jj'}^{s, {\rm R}} + i\mathcal{N}_{jj'}^{s, {\rm I}} &= \sum_{l l' m} \sum_{k k' p} \left[\hat{\zeta}_{l l'}^{m, {\rm R}} + i \hat{\zeta}_{l l'}^{m, {\rm I}} \right]\left[\mathcal{R}_{k k'}^{p,{\rm R}} + i \mathcal{R}_{k k'}^{p,{\rm I}}\right]\nonumber\\ &\times\mathcal{G}_{l k j}^{m p -s} \mathcal{G}_{l' k' j'}^{-m -p s}
\label{eqn:edge_complex}
\end{align}
where superscript ${\rm R}$ denotes the real part and superscript ${\rm I}$ the imaginary part. Inverting equation (\ref{eqn:zeta_symm_defn}) we may replace the real part of each coefficient in equation (\ref{eqn:edge_complex}) in terms of the symmetrized coefficient, yielding
\begin{align}
\frac{1}{2-\delta^{\rm K}_{s0}}\bar{\mathcal{N}}_{jj'}^{|s|} +& i\mathcal{N}_{jj'}^{s, {\rm I}} = \sum_{l l' m} \sum_{k k' p} \left[\frac{1}{2-\delta^{\rm K}_{m0}}  \hat{\bar{\zeta}}_{l l'}^{|m|} + i \hat{\zeta}_{l l'}^{m, {\rm I}} \right]\nonumber\\
&\times\left[\frac{1}{2-\delta^{\rm K}_{p0}}   
\bar{\mathcal{R}}_{k k'}^{|p|} + i \mathcal{R}_{k k'}^{p,{\rm I}}\right]\mathcal{G}_{l k j}^{m p -s} \mathcal{G}_{l' k' j'}^{-m -p s};
\label{eqn:edge_complex_subst}
\end{align}
the absolute value signs on the spins are necessary because their range includes negative values here but the symmetrized coefficients are defined only for positive or zero spins.  

We seek $\hat{\bar{\zeta}}_{ll'}^{|m|}$ and would like to estimate it solely in terms of $\bar{\mathcal{N}}_{jj'}^{|s|}$ and $\bar{\mathcal{R}}_{kk'}^{|p|}$. We see that if we set the imaginary part of the randoms' unsymmetrized coefficient to zero (indeed, we do not wish to track it in our algorithm), this ensures the real part of the righthand side of equation (\ref{eqn:edge_complex_subst}) involves only our desired $\hat{\bar{\zeta}}_{ll'}^{|m|}$. In contrast, the imaginary part remaining on the righthand side will involve $\hat{\zeta}_{ll'}^{m,{\rm I}}\bar{\mathcal{R}}_{kk'}^{|p|}$ and thus has no information about $\hat{\bar{\zeta}}_{ll'}^{|m|}$. We may now take the real part of the whole equation, yielding
\begin{align}
\frac{1}{2-\delta^{\rm K}_{s0}}\bar{\mathcal{N}}_{jj'}^{|s|} =& \sum_{l l' m} \sum_{k k' p} \frac{1}{2-\delta^{\rm K}_{m0}}  \hat{\bar{\zeta}}_{l l'}^{|m|} \nonumber\\
&\times\frac{1}{2-\delta^{\rm K}_{p0}}   
\bar{\mathcal{R}}_{k k'}^{|p|} \mathcal{G}_{l k j}^{m p -s} \mathcal{G}_{l' k' j'}^{-m -p s};
\label{eqn:edge_corrxn_symm}
\end{align}

Following \cite{SE_3pt_alg}, we now divide both sides of equation (\ref{eqn:edge_corrxn_symm}) through by $\bar{\mathcal{R}}_{00}^0$ and separate off the $kk'p =000$ term on the right hand side, finding
\begin{align}
\frac{\bar{\mathcal{N}}_{j j'}^{|s|}}{(2-\delta^{\rm K}_{s0})\bar{\mathcal{R}}_{00}^0} =& \frac{1}{2-\delta^{\rm K}_{s0}}\hat{\bar{\zeta}}_{j j'}^{|s|}  + \sum_{l l' m} \frac{1}{2-\delta^{\rm K}_{m0}}\hat{\bar{\zeta}}_{l l'}^{|m|}\nonumber\\
&\times\sum_{k k' p \neq 000}\frac{1}{2-\delta^{\rm K}_{p0}}\bar{f}_{k k'}^{|p|} \mathcal{G}_{lkj}^{m p -s} \mathcal{G}_{l' k' j'}^{-m -p s}
\label{eqn:edge_corrxn_simp}
\end{align}
where $\bar{f}_{k k'}^{|p|} \equiv \bar{\mathcal{R}}_{k k'}^{|p|}/\bar{\mathcal{R}}_{00}^0$. 

In the isotropic case, $l' = l$ and $\bar{\zeta}_{jj'}^{|s|}$ and $\bar{\zeta}_{ll'}^{|m|}$ become spin-independent. Thus in this limit we can sum over all spins to estimate the isotropic 3PCF $\zeta_l$ of \cite{SE_3pt_alg}. The only factors inside the sums over $m$ and $s$ on the right-hand side then are 3j-symbols, and invoking the orthogonality identity \citet{NIST_DLMF} 34.3.16 we find that $k=k'$. We then sum the resultant $f_{kk}^p$ over $p$ weighted by the Gaunt integrals. This sum can be identified with the edge-correction factors $f_l'$ of \cite{SE_3pt_alg}, showing that only isotropic edge-correction factors enter an estimate of the isotropic 3PCF. 

We now define an edge correction matrix ${\bf M}$ with elements
\begin{align}
M_{l l' m}^{ j j' s} = \sum_{k k' \neq 00}\bar{f}_{k k'}^{|p|} \mathcal{G}_{lkj}^{m p -s} \mathcal{G}_{l' k' j'}^{-m -p s},\;\;\;p=s-m,
\label{eqn:edge_matrix}
\end{align}
using that the Gaunt symbol forces $m+p-s=0$ to set $p$. In terms of this matrix, the edge-correction equation (\ref{eqn:edge_corrxn_simp}) becomes
\begin{align}
\frac{\vec{\bar{\mathcal{N}}}}{\bar{\mathcal{R}}_{00}^0}  =\left({\bf I} + {\bf M} \right)\vec{\hat{\bar{\zeta}}}
\end{align}
where $\vec{\bar{\mathcal{N}}}$ is a vector of $\bar{\mathcal{N}}_{j j'}^{|s|}$, the double spherical harmonic moments of the counts, $\vec{\hat{\bar{\zeta}}}$ is a vector of the double spherical harmonic moments $\hat{\bar{\zeta}}_{ll'}^{|m|}$, and ${\bf I}$ is the identity matrix. Given the coefficients $\bar{f}_{k k'}^{|p|}$, which encode the impact of the survey geometry on the randoms, and $\bar{\mathcal{R}}_{00}^0$, the overall normalization of the random triple count, this equation can be solved by matrix inversion for the desired $\vec{\hat{\bar{\zeta}}}$.

We note that there are two approximations involved in solving the edge correction equation here. First, to obtain a given matrix element, one formally requires all values of the random coefficients $\mathcal{R}$ to perform the sum in equation (\ref{eqn:edge_matrix}). However, in practice, as the next subsection will illustrate, these values fall so quickly with rising $k$ and $k'$ that we expect the full matrix element is very well-approximated by truncating at some $k$ and $k'$ of order the maximal $l$ and $l'$ to which an anisotropic 3PCF measurement is desired.

Second, the edge correction matrix is formally infinite in dimension, as at fixed $l$ and $l'$ all $j$ and $j'$ can enter the correction. Thus one should also measure an infinite set of coefficients $\bar{\mathcal{N}}_{jj'}^{|s|}$ of the $D-R$ field. However, again, in practice, truncating this set at roughly the maximal $l$ or $l'$ desired for the 3PCF measurement should be sufficiently accurate to correct for the survey geometry. This point is detailed further in \cite{SE_3pt_alg} for the isotropic 3PCF, and the same outcome is expected for the anisotropic 3PCF. 

Summarizing, the rapid fall-off of the edge correction factors means that truncating the sums required for each matrix element is a good approximation, and the expected near-diagonality of the matrix means that the truncation of the matrix itself will not greatly affect the inverse.

\subsection{A toy model for the edge correction factors}
\label{subsec:edge_corrxn_toy}
Here we investigate a toy model of the survey geometry to gain intuition for the values of the edge correction factors. We consider a planar survey boundary perpendicular to the line of sight (i.e. the $z$ axis). For a given primary galaxy, the sphere around the primary over which we compute the 3PCF will impinge upon the survey boundary if the primary is less than $R_{\rm max}$ away from it. These galaxies will be only a small fraction of the total for realistic survey geometries, so the edge-correction factors we estimate from this model will be diluted by the bulk of the survey where a primary's sphere is fully contained, and we quantify this dilution at the end of the section.

For now we focus on primaries whose surrounding sphere does impinge on the survey boundary. We take it that a given primary is a distance $z$ away from the survey boundary. This model was explored in \cite{SE_3pt_alg} but only the isotropic edge correction factors $f_l = 4\pi/(2l+1)\sum_m f_{ll}^m$ were computed there, as they were the only factors relevant for correcting the isotropic 3PCF. Here, we seek the more general
\begin{align}
f^0_{kk'} \equiv \frac{\mathcal{R}_{kk'}^0}{\mathcal{R}_{00}^0}=\frac{\left<a_{k0}a_{k'0}^*\right>}{a_{00}^2}
\label{eqn:f_est}
\end{align}
and angle brackets represent azimuthal averaging. We note that the azimuthal symmetry of our toy model means the only non-zero edge correction factors are those with zero spin.

For a secondary galaxy $R_{\rm max}$ distant from the primary , there will be a critical angle $\mu_c=z/R_{\rm max}$ for which it  is outside the survey for larger $\mu$. As shown in \cite{SE_3pt_alg} equation (35), the spherical harmonic coefficients are then
\begin{align}
a_{l0}=\sqrt{\frac{\pi}{2l+1}}[\mathcal{L}_{l+1}(\mu_c) - \mathcal{L}_{l-1}(\mu_c)]
\end{align}
for $l\geq 1$. We note that there is no dependence on the triangle side length because these coefficients represent a random density field. We also note that since the only relevant coefficients for this toy model are at spin zero, the symmetrization discussed in \S\ref{sec:basis} does not affect our results here.

To obtain $f^0_{kk'}$, we now form the product of the spherical harmonic coefficients equation (\ref{eqn:f_est}) demands and average over $\mu_c$, as
\begin{align}
&\left<a_{k0}a_{k'0}^*\right> =\frac{\pi}{\sqrt{(2k+1)(2k'+1)}} \nonumber\\
&\times\int_0^1 d\mu_c \bigg[ \mathcal{L}_{k+1}(\mu_c) \mathcal{L}_{k'+1}(\mu_c)  - \mathcal{L}_{k-1}(\mu_c) \mathcal{L}_{k'+1}(\mu_c) \nonumber\\
&- \mathcal{L}_{k+1}(\mu_c)\mathcal{L}_{k' -1}(\mu_c)+ \mathcal{L}_{k-1}(\mu_c)\mathcal{L}_{k'-1}(\mu_c)\bigg].
\label{eqn:toy_model_integral}
\end{align}
We may perform this integral by using the linearization formula for Legendre polynomials given in \nameref{sec:appendix_A} to convert the products of Legendres into sums over a single Legendre and then integrating using the recursion relation also given in \nameref{sec:appendix_A}. We find
\begin{align}
&\left<a_{k0}a_{k'0}^*\right> =\frac{\pi}{\sqrt{(2k+1)(2k'+1)}} \sum_J \bigg[\left(\begin{array}{ccc}
k+1 & k'+1 & J\\
0 & 0 & 0
\end{array}\right)^2\nonumber\\
&-\left(\begin{array}{ccc}
k-1 & k'+1 & J\\
0 & 0 & 0
\end{array}\right)^2
-\left(\begin{array}{ccc}
k+1 & k'-1 & J\\
0 & 0 & 0
\end{array}\right)^2\nonumber\\
&+\left(\begin{array}{ccc}
k-1 & k'-1 & J\\
0 & 0 & 0
\end{array}\right)^2
\bigg]\left[\mathcal{L}_{J-1}(0)-\mathcal{L}_{J+1}(0)\right].
\label{eqn:muc_avg}
\end{align}
To simplify we observed that $\mathcal{L}_J(1)=1$ for all $J$ and so the terms from the upper bound of the integral (\ref{eqn:toy_model_integral}) vanish.\footnote{This sum of four 3j-symbols can be further simplified using recursions 8.6.4.21 and 8.6.4.23 in \cite{VMK06} to a pre-factor times a single 3j-symbol if desired.} We note that $J$ must be odd or the difference of Legendres in the last line of equation (\ref{eqn:muc_avg}) vanishes, as $\mathcal{L}_n(0) = 0$ for $n$ odd by parity.

We pause to observe that in the limit where $k' \simeq  k$ and $k\to \infty$, $J$ is constrained to be small ($J \leq |k-k'|$).  We can then use the square of the 3j-symbol's asymptotic, which is $\simeq d^2/(2k+1)$, where $d$ is a Wigner matrix and $|d^2| \leq 1$. Combining this with the pre-factor $\propto 1/(2k+1)$, we see that the limit of the isotropic edge correction factors scales as $1/(2k+1)^2$, going to zero as $k$ grows. This roughly recovers the limiting behavior of equation (38) for the isotropic edge correction factor in \cite{SE_3pt_alg}.

From explicit computation in \cite{SE_3pt_alg} we have that $\mathcal{R}_{00}^0=(a_{00})^2 = (7\pi)/3$, so we may form the ratio $f_{kk'}^0$ of equation (\ref{eqn:f_est}).\footnote{We use the first line of equation (35) of \cite{SE_3pt_alg} with $l=0$ and then average over $\mu_{\rm c}$.} We compute a number of values to show that the edge correction factors are indeed small: $f_{10}^0 = -9.28\%,f_{11}^0 =17.14\%,  f_{20}^0 =-6.39\%, f_{12}^0 =6.92\%, f_{30}^0 =-2.36\%,  f_{31}^0= -1.87\%, f_{22}^0 =4.08\% ,f_{41}^0 =  -2.90\%, f_{32}^0=0.66\%,f_{50}^0  =0.37\%, f_{42}^0 =-0.91\%, f_{33}^0 =1.90\%$. The factors where $k=k'$ correspond to the $f_k$ of \cite{SE_3pt_alg}, and we recover the same values here as in that work. We point out that the edge correction factors in this toy model do not satisfy the constraint that $k+k'$ is even; our survey boundary is a plane perpendicular to the $z$-axis and so the random clustering is not invariant under parity.

In an SDSS BOSS-like geometry, only $20\%$ of galaxies lie within $R_{\rm max} = 200\Mpc$ of a survey boundary, so these factors should be further scaled down by roughly a factor of $5$. On the other hand, the true survey geometry is more complicated than our simple planar toy model, so the edge correction factors estimated above should be taken only as a rough guide.

Finally, we note that the calculation of this section could also be applied to estimate the edge-correction factors' values for an angular survey boundary. For this situation the boundary has rotated $90^{\degree}$ from the planar redshift-boundary explored above, or equivalently we can consider that the $z$-axis of coordinates has rotated by this amount. Thus we can compute the harmonic coefficients of this angular boundary by rotating all of the redshift-boundary coefficients prior to squaring and taking their expectation value over $\mu_{\rm c}$. This rotation can be accomplished using Wigner D-matrices. Since the D-matrices are rotations, and thus unitary, we expect that an angular boundary would not lead to significantly different amplitudes of the edge correction factors from those in the redshift-boundary case; thus our overall claim that these factors should be small still holds.

\section{Covariance of the anisotropic 3PCF}
\label{sec:covar}
\subsection{Adapting the real-space calculation to redshift-space}
Here we adapt the isotropic 3PCF covariance calculation of \cite{SE_3pt_alg} for the anisotropic case. We use a tilde to denote a Fourier space quantity, use a negative exponential for inverse Fourier Transforms, and always use  $d^3\vec{k}/(2\pi)^3$ when going from Fourier space to configuration space.  Adapting \cite{SE_3pt_alg} equation (45) by replacing the real-space density perturbation $\delta$ by its redshift-space analog $\delta_s$, we see that in equation (47) of that work the Fourier-space density perturbations $\tilde{\delta}$ can be replaced by their redshift-space analogs $\tilde{\delta}_s(\vec{k})$.

Following \cite{SE_3pt_alg}, we use Wick's Theorem to contract the six Fourier space density fields implied by the 3PCF covariance. We then adapt equation (49) of that work by replacing the isotropic power spectra with their multipole analogs,
\begin{align}
P(k;\mu) = \sum_{l_k} P_{l_k}(k)\mathcal{L}_{l_k}(\mu)
\end{align}
We emphasize that this expansion for the anisotropic power spectrum is fully general, though in the flat-sky (``Kaiser'') approximation and under linear theory the series reduces further to have terms only at $l=0,2,$ and $4$, as we will discuss further in \S\ref{subsec:kaiser_covar}.  Following through to the analogs of \cite{SE_3pt_alg} equations (50) and (51), we find
\begin{align}
&{\rm Cov} \equiv \left< \hat{\zeta}(\vec{r}_1,\vec{r}_2)\hat{\zeta}(\vec{r}_1',\vec{r}_2') \right>=\frac{1}{V}
\int \frac{d^3\vec{q}\;d^3\vec{p}\;d^3\vec{k}}{(2\pi)^9}\nonumber\\
&\times \sum_{l_q l_p l_k}P_{l_q}(q) P_{l_p}(p) P_{l_k}(k)\mathcal{L}_{l_q}(\mu_q)\mathcal{L}_{l_p}(\mu_p) \mathcal{L}_{l_k}(\mu_k)\nonumber\\
&\times (2\pi)^3\delta_{D}^{[3]}(\vec{q}+\vec{p}+\vec{k}) e^{-i\left[\vec{q}\cdot\vec{r}_{1}+\vec{p}\cdot\vec{r}_{2}\right]}\nonumber\\
&\times\bigg\{ e^{-i\left[\vec{q}\cdot\vec{r}_{1}'+\vec{p}\cdot\vec{r}_{2}'\right]}+e^{-i\left[\vec{p}\cdot\vec{r}_{1}'+\vec{q}\cdot\vec{r}_{2}'\right]}+e^{-i\left[\vec{k}\cdot\vec{r}_{1}'+\vec{p}\cdot\vec{r}_{2}'\right]}\nonumber\\
&+e^{-i\left[\vec{p}\cdot\vec{r}_{1}'+\vec{k}\cdot\vec{r}_{2}'\right]}+e^{-i\left[\vec{k}\cdot\vec{r}_{1}'+\vec{q}\cdot\vec{r}_{2}'\right]}+e^{-i\left[\vec{q}\cdot\vec{r}_{1}'+\vec{k}\cdot\vec{r}_{2}'\right]}\bigg\}.
\label{eqn:full_covar}
\end{align}

\subsection{Projection onto our basis}
We now show how to obtain the covariance of our symmetrized harmonic coefficients for the 3PCF from the full covariance. We will begin by projecting the covariance onto the full spherical harmonic basis including negative spins and at the end of the calculation show how to obtain the covariance of the symmetrized coefficients from these results. 

The covariance projected onto our full spherical harmonic basis is
\begin{align}
&{\rm Cov}_{l_1 l_2 m, l_1' l_2' m'}(r_1, r_2;r_1', r_2') =\frac{\sqrt{(2l_1+1)(2l_2+1)(2l_1'+1)(2l_2'+1)}}{(4\pi)^2} \nonumber\\
&\times\int d\Omega_{r_1}d\Omega_{r_2}d\Omega_{r'_1}d\Omega_{r'_2}Y^*_{l_1 m}(\hat{r}_1)Y_{l_2 m}(\hat{r}_2) \nonumber\\
&\times Y_{l_1'm'}^*(\hat{r}'_1)Y_{l_2' m'}(\hat{r}_2') {\rm Cov}(\vec{r}_1,\vec{r}_2;\vec{r}_1',\vec{r}_2').
\end{align}
The pre-factor comes from two copies of the inverse of the pre-factor in equation (\ref{eqn:full_3pcf_est}), as we wish to extract the covariance of the coefficients $\zeta_{ll'}^m$.
Noticing that the only dependence on real-space variables in the full covariance is in the exponentials, we require the projection integrals
\begin{align}
&\int d\Omega_r \;e^{-i\vec{k}\cdot \vec{r}} Y^*_ {lm}(\hat{r}) \nonumber\\
&=(4\pi)(-i)^l j_l(kr) Y^*_{lm}(\hat{k})
\label{eqn:projxn_int_def}
\end{align}
and
\begin{align}
&\int d\Omega_r \;e^{-i\vec{k}\cdot \vec{r}} Y_ {lm}(\hat{r})\nonumber\\
&=(4\pi)(-i)^l j_l(kr) Y_{lm}(\hat{k})
\end{align}
where we used the plane wave expansion (\citealt{AWH13} equation 16.61) to expand the exponential into spherical Bessel functions and spherical harmonics and then invoked orthogonality. 

The projection of all of the exponentials in equation (\ref{eqn:full_covar}) is then
\begin{align}
&\mathcal{E}_{l_1 l_2 m, l_1' l_2' m' }\equiv \sqrt{(2l_1+1)(2l_2+1)(2l_1'+1)(2l_2'+1)}\nonumber\\
& \times (-i)^{l_1 +l_2 +l_1' +l_2'}\mathcal{J}_{l_1l_2}(qr_1; pr_2) Y^*_{l_1 m}(\hat{q})Y_{l_2 m}(\hat{p}) \nonumber\\
&\times \bigg\{\mathcal{J}_{l_1' l_2'}(qr_1'; pr_2') Y^*_{l_1' m'}(\hat{q})Y_{l_2' m'}(\hat{p})\nonumber\\
&+ \mathcal{J}_{l_1' l_2'}(pr_1'; qr_2') Y^*_{l_1' m'}(\hat{p})Y_{l_2' m'}(\hat{q})\nonumber\\
&+\mathcal{J}_{l_1' l_2'}(kr_1'; pr_2') Y^*_{l_1' m'}(\hat{k})Y_{l_2' m'}(\hat{p})\nonumber\\
&+ \mathcal{J}_{l_1' l_2'}(pr_1'; kr_2') Y^*_{l_1' m'}(\hat{p})Y_{l_2' m'}(\hat{k})\nonumber\\
&+\mathcal{J}_{l_1' l_2'}(kr_1'; qr_2') Y^*_{l_1' m'}(\hat{k})Y_{l_2' m'}(\hat{q})\nonumber\\
&+\mathcal{J}_{l_1' l_2'}(qr_1'; kr_2') Y^*_{l_1' m'}(\hat{q})Y_{l_2' m'}(\hat{k})\bigg\}, 
\label{eqn:epsilon}
\end{align}
where we have used equation (\ref{eqn:projxn_int_def}) and simplified. We have defined
\begin{align}
\mathcal{J}_{l_1 l_2}(qr_1;pr_2) \equiv j_{l_1}(qr_1) j_{l_2}(pr_2).
\end{align}
The symmetry structure of these terms can be easily checked: within the curly brackets, the first and second term are equal under $q \leftrightarrow p$, the third and fourth under $k \leftrightarrow p$, and the fifth and sixth under $k \leftrightarrow q$.

Returning to equation (\ref{eqn:full_covar}) and using \cite{SE_3pt_alg} equation (58) to expand the Dirac delta function into an integral over plane waves and thence into spherical harmonics and spherical Bessel functions, and the spherical harmonic addition theorem to expand the Legendre polynomials (\citealt{AWH13} equation 16.57), the projected covariance is
\begin{align}
&{\rm Cov}_{l_1 l_2 m,l_1' l_2' m'}(r_1, r_2;r_1', r_2') =\nonumber\\
&\frac{1}{V} \int \frac{d^3\vec{q}\; d^3\vec{p}\; d^3\vec{k} }{(2\pi)^9} \sum_{l_q l_p l_k}P_{l_q}(q) P_{l_p}(p)P_{l_k}(k)\nonumber\\
&\times \frac{(4\pi)^3}{(2l_q+1)(2l_p+1)(2l_k+1)}\sqrt{\frac{(2l_q+1)(2l_p+1)(2l_k+1)}{(4\pi)^3}} \nonumber\\
&\times Y_{l_q 0}(\hat{q})Y_{l_p 0}(\hat{p}) Y_{l_k 0} (\hat{k})\nonumber\\
&\times (4\pi)^3\sum_{J_1 J_2 J_3}\sum_{S_1 S_2 S_3} \mathcal{D}_{J_1 J_2 J_3}\mathcal{C}_{J_1 J_2 J_3}\mathcal{R}_{J_1 J_2 J_3}(q, p, k)\nonumber\\
&\times 
\left(\begin{array}{ccc}
J_1 & J_2 & J_3\\
S_1 & S_2 & S_3
\end{array}\right)  Y_{J_1 S_1}^*(\hat{q})Y_{J_2 S_2}^*(\hat{p})Y_{J_3 S_3}^*(\hat{k}) \nonumber\\
&\times\mathcal{E}_{l_1 l_2 m, l_1' l_2' m'}.
\label{eqn:proj_covar}
\end{align}
The first factor in the third line stems from the spherical harmonic addition theorem applied to the three Legendre polynomials entering the power spectrum multipoles. The second factor comes from the spin-zero values of the spherical harmonics of the line of sight (taken to be the $z$-axis); these are the only spherical harmonics entering the power spectrum multipole decomposition. The fourth line comes from the spherical harmonic expansion of the power spectrum's multipole moments, and the fifth and sixth lines come from the Dirac delta function's expansion into spherical harmonics, with 
\begin{align}
&\mathcal{D}_{J_1 J_2 J_3} \equiv i^{J_1 + J_2 +J_3}\nonumber\\
&\mathcal{C}_{J_1 J_2 J_3} \equiv \sqrt{\frac{(2J_1 +1)(2J_2+1)(2J_3+1)}{4\pi}}\nonumber\\
&\mathcal{R}_{J_1 J_2 J_3}(q, p, k)\equiv \int r^2 dr j_{J_1}(qr) j_{J_2}(pr) j_{J_3}(kr).
\label{eqn:delta_fn_defs}
\end{align}

For each Fourier space unit vector in equation (\ref{eqn:proj_covar}), there is one spherical harmonic contributed by the expansion of the Legendre polynomial from the power spectrum multipoles, another from the Dirac delta function, and either zero, one or two from $\mathcal{E}_{{\rm proj},l_1 l_2 m, l_1' l_2' m'}$. Thus when we perform the integrations over angles $d\Omega_q d\Omega_p d\Omega_k$ we will have integrals of two, three, or four spherical harmonics. 

In particular, the first two terms in the curly brackets in equation (\ref{eqn:epsilon}) will lead to four spherical harmonics in $\hat{p}$ and $\hat{q}$ and two in $\hat{k}$; the final four terms in the curly brackets will give four in $\hat{p}$ and three in $\hat{q}$ or vice versa and always three in $\hat{k}$. We thus work in terms of the Gaunt integral $\mathcal{G}$ of three spherical harmonics (all unconjugated) and a generalized Gaunt integral $\mathcal{H}$ of four spherical harmonics (again all unconjugated). Explicit expressions for $\mathcal{G}$ and $\mathcal{H}$ are given in \nameref{sec:appendix_A}. Performing the angular integrals, the covariance becomes
\begin{align}
&{\rm Cov}_{l_1 l_2 m, l_1' l_2' m'}(r_1, r_2;r_1', r_2') =(-i)^{l_1 + l_2 +l_1' + l_2'}\nonumber\\
& \times \frac{1}{V} \int \frac{k^2dk\; p^2 dp\; q^2 dq }{(2\pi^2)^3} \sum_{l_q l_p l_k}P_{l_q}(q) P_{l_p}(p)P_{l_k}(k)\nonumber\\
&\times (4\pi)^{3/2}\sqrt{\frac{(2l_1+1)(2l_2+1)(2l_1'+1)(2l_2'+1)}{(2l_q+1)(2l_p+1)(2l_k+1)}}\nonumber\\
&\times\sum_{J_1 J_2 J_3}\sum_{S_1 S_2 S_3} \mathcal{D}_{J_1 J_2 J_3}\mathcal{C}_{J_1 J_2 J_3}\mathcal{R}_{J_1 J_2 J_3}(q, p, k)\nonumber\\
&\times \left(\begin{array}{ccc}
J_1 & J_2 & J_3\\
0 & 0 & 0
\end{array}\right)
\left(\begin{array}{ccc}
J_1 & J_2 & J_3\\
S_1 & S_2 & S_3
\end{array}\right)\mathcal{J}_{l_1 l_2}(qr_1,pr_2)\nonumber\\
&\times (-1)^{m+m'}\bigg\{\delta^K_{J_3 l_k}\delta^K_{S_3 0} \bigg[\mathcal{J}_{l_1' l_2'}(q,p)\mathcal{H}_{l_q J_1 l_1 l_1'}^{0 -S_1 -m -m'}\mathcal{H}_{l_p J_2 l_2 l_2'}^{0 S_1 m m'}\nonumber\\
&+\mathcal{J}_{l_1' l_2'}(p,q)\mathcal{H}_{l_q J_1 l_1 l_2'}^{0 -S_1 -m m'}\mathcal{H}_{l_p J_2 l_2 l_1'}^{0 S_1 m -m'}\bigg]\nonumber\\
&+\mathcal{J}_{l_1' l_2'}(k,p)\mathcal{G}_{l_q J_1 l_1}^{0 -S_1 -m} \mathcal{H}_{l_p J_2 l_2 l_1'}^{0 -S_2 m m'} \mathcal{G}_{l_k J_3 l_1'}^{0 -S_3 -m'}\nonumber\\
&+\mathcal{J}_{l_1' l_2'}(p,k)\mathcal{G}_{l_q J_1 l_1}^{0 -S_1 -m} \mathcal{H}_{l_p J_2 l_2 l_1'}^{0 -S_2 m -m'} \mathcal{G}_{l_k J_3 l_2'}^{0 -S_3 m'}\nonumber\\
&+\mathcal{J}_{l_1' l_2'}(k,q) \mathcal{H}_{l_q J_1 l_1 l_2'}^{0 -S_1 -m m'}  \mathcal{G}_{l_p J_2 l_2}^{0 -S_2 m} \mathcal{G}_{l_k J_3 l_1'}^{0 -S_3 -m'}\nonumber\\
&+\mathcal{J}_{l_1' l_2'}(q,k) \mathcal{H}_{l_q J_1 l_1 l_1'}^{0 -S_1 -m -m'}  \mathcal{G}_{l_p J_2 l_2}^{0 -S_2 m} \mathcal{G}_{l_k J_3 l_2'}^{0 -S_3 m'}\bigg\}
\label{eqn:full_covar_finished}
\end{align}
We emphasize that we never do arithmetic in the indices of $\mathcal{G}$ or $\mathcal{H}$, so a minus sign on an index should always be interpreted to mean the negative of the indicated variable. Relative to equations (\ref{eqn:epsilon}) and (\ref{eqn:proj_covar}), we used the identity that $Y_{lm}^*=(-1)^mY_{l -m}$ to ensure all spherical harmonics were unconjugated prior to integration, and we then used that $S_1 + S_2 + S_3=0$ to simplify (required by the 3j-symbol). In the first and second term this condition means that $S_1 + S_2 = 0$ because the Kronecker delta sets $S_3 = 0$. For Gaunt integrals, the spins must sum to zero, and for $\mathcal{H}$ the first two spins must have sum equal and opposite to the sum of the last two spins.\footnote{This will always be true for two pairs of spins but the fact that here the pairing is first two and last two is specific to how we chose to couple spherical harmonics when evaluating $\mathcal{H}$.} These rules set $S_1, S_2$, and $S_3$ in equation (\ref{eqn:full_covar_finished}). In fact, they over-constrain it, and so ensuring both rules can be satisfied for each term acts as a consistency check.

We now show that this expression requires only finite sums over angular momenta. The Gaunt integrals $\mathcal{G}$ require that the total momenta form a closed triangle (``closure condition''), and the spins in turn are bounded by the total momenta. Similarly, the generalized Gaunt integrals $\mathcal{H}$ require that the total momenta form a closed quadrilateral, and the spins are again bounded by the total momenta. Of the momenta, only the $J_i$ are free, and these appear at most once in each $\mathcal{G}$ or $\mathcal{H}$; thus the other, constrained sides along with the closure condition will bound the $J_i$ so that all the sums over momenta are finite. In particular, in the Kaiser limit for the anisotropic power spectrum,  $l_q, l_p$, and $l_k$ take on values $0,2,$ and $4$. The other momenta, $l_1,l_2,l_1'$, and $l_2'$, are fixed by the covariance matrix element desired. 

\subsection{Reduction to integrals of the power spectrum's multipole moments}
While equation (\ref{eqn:full_covar_finished}) appears involved, it is actually a considerable simplification of the covariance as regards calculation. Recalling that the $\mathcal{J}$ are simply products of spherical Bessel functions with arguments given by the Fourier space magnitude noted ($p,q,$ or $k$) times a configuration space side length, we will show that equation (\ref{eqn:full_covar_finished}) enables computation of the covariance with 1-D and 2-D integral transforms. In particular, all of the wave-vector magnitude dependence in the projected covariance can be written in terms of $f$-tensors
\begin{align}
&f_{nm}^{l}(r; r_i) = \int \frac{k^2 dk}{2\pi^2}P_{l}(k)j_{n}(kr)j_m(kr_i)\nonumber\\
&f_{nmj}^{l}(r;r_i, r_j') = \int \frac{k^2 dk}{2\pi^2} P_{l}(k) j_n(kr)j_m(kr_i)j_j(kr_j').
\end{align}
Given that the subscripted variables will be binned into of order $\Nbins = 10$ bins and the spherical Bessel functions of these arguments replaced with their bin-averaged analogs, these tensors need be computed only on a fine grid in $r$, not in the $r_i$ or the $r'_i$. Recall that $r$ is a dummy variable to be integrated over to enforce the Dirac delta function constraint.

We now show how these $f$-tensors emerge. Focusing solely on the dependence on the Fourier-space wave-vectors' magnitudes, the first term in equation (\ref{eqn:full_covar_finished}) scales as
\begin{align}
&\int \frac{k^2 dk}{2\pi^2}P_{l_k}(k)j_{l_k}(kr)\int\frac{q^2 dq}{2\pi^2}P_{l_q}(q)j_{J_1}(qr)j_{l_1}(qr_1)j_{l_1'}(qr_1')\nonumber\\
&\times \int\frac{p^2 dp}{2\pi^2}P_{l_p}(p) j_{J_2}(pr)j_{l_2}(pr_2)j_{l_2'}(pr_2'),
\label{eqn:first_term}
\end{align}
and the first factor further simplifies to
\begin{align}
\int \frac{k^2 dk}{2\pi^2}P_{l_k}(k)j_{l_k}(kr) =\xi_{l_k}(r),
\end{align}
where $\xi_{l_k}(r)$ is the $l_k^{th}$ multipole moment of the anisotropic 2PCF. The second term in equation (\ref{eqn:full_covar_finished}) has the same form as the first term but with $r_1'\leftrightarrow r_2'$.

Again focusing solely on the dependence on wave-vector magnitudes, the third term in equation (\ref{eqn:full_covar_finished}) scales as
\begin{align}
&\int \frac{q^2 dq}{2\pi^2} P_{l_q}(q) j_{J_1}(qr)j_{l_1}(qr_1) \int\frac{k^2 dk}{2\pi^2} P_{l_k}(k) j_{J_3}(kr)j_{l_1'}(kr_1')\nonumber\\
&\times \int\frac{p^2 dp}{2\pi^2}P_{l_p}(p)j_{J_2}(pr)j_{L_2}(pr_1)j_{l_2'}(pr_2').
\label{eqn:third_term}
\end{align}
The fourth term is given by equation (\ref{eqn:third_term}) with $r_1'\leftrightarrow r_2'$.

Finally, the wave-vector magnitude dependence of the fifth term in equation (\ref{eqn:full_covar_finished}) scales as
\begin{align}
&\int \frac{p^2 dp}{2\pi^2} P_{l_p}(p) j_{J_2}(pr)j_{l_2}(pr_2) \int \frac{k^2 dk}{2\pi^2} P_{l_k}(k)j_{J_3}(kr) j_{l_1'}(kr_1')\nonumber\\
&\times \int \frac{q^2 dq}{2\pi^2}P_{l_q}(q) j_{J_1}(qr)j_{l_1}(qr_1)j_{l_2'}(qr_2'),
\label{eqn:fifth_term}
\end{align}
with the sixth term given by switching $r_1'\leftrightarrow r_2'$ above.

\subsection{Elimination of the Delta function's spins}
We observe that the $f$-tensors depend only on the total angular momenta, not the spins. Moreover, as noted earlier, the integrals $\mathcal{G}$ and $\mathcal{H}$ determine $S_1, S_2$, and $S_3$ in equation (\ref{eqn:full_covar_finished}), so we may eliminate the sum over $S_1, S_2$ and $S_3$. We would expect one can resum without these spins because they enter only the Dirac delta function and it is in total spin-independent. Thus new weights $w_i,i=1-6$ for each term can be defined that depend only on the total angular momenta and the free spins $m$ and $m'$. All weights have the same argument: $w_i=w_i(l_q l_p l_k; l_1 l_2; l_1' l_2'; m m')$; below we suppress it for brevity. The weights are
\begin{align}
&w_1 =\mathcal{H}_{l_q J_1 l_1 l_1'}^{0 -S_1 -m -m'}\mathcal{H}_{l_p J_2 l_2 l_2'}^{0 S_1 m m'},\;\;\;S_1=-m-m'\nonumber\\
&w_2= \mathcal{H}_{l_q J_1 l_1 l_2'}^{0 -S_1 -m m'}\mathcal{H}^{0 S_1 m -m'}_{l_p J_2 l_2 l_1'},\;\;\;S_1=m'-m\nonumber\\
&w_3= \mathcal{G}_{l_q J_1 l_1}^{0 m -m} \mathcal{H}_{l_p J_2 l_2 l_1'}^{0 -S_2 m m'} \mathcal{G}_{l_k J_3 l_1'}^{0 m' -m'},\;\;\;S_2=m+m'\nonumber\\
&w_4=  \mathcal{G}_{l_q J_1 l_1}^{0 m -m} \mathcal{H}_{l_p J_2 l_2 l_1'}^{0 -S_2 m -m'} \mathcal{G}_{l_k J_3 l_2'}^{0 -m' m'},\;\;\;S_2=m - m'\nonumber\\
&w_5 = \mathcal{H}_{l_q J_1 l_1 l_2'}^{0 -S_1 -m m'}  \mathcal{G}_{l_p J_2 l_2}^{0 -m m} \mathcal{G}_{l_k J_3 l_1'}^{0 m' -m'},\;\;\;S_1=m'-m\nonumber\\
&w_6 = \mathcal{H}_{l_q J_1 l_1 l_1'}^{0 -S_1 -m -m'}  \mathcal{G}_{l_p J_2 l_2}^{0 -m m} \mathcal{G}_{l_k J_3 l_2'}^{0 -m' m'},\;\;\;S_1=-m-m'.
\label{eqn:wts}
\end{align}
We have explicitly replaced the spins $S_i$ that can be written simply as $\pm m$ or $\pm m'$, but indicated above at right the values of spins written as sums or differences of $m$ and $m'$ to avoid ambiguous superscripts.

Written using these weights (with arguments suppressed), the covariance becomes
\begin{align}
&{\rm Cov}_{l_1 l_2m,l_1' l_2'm'}(r_1, r_2;r_1', r_2')=\frac{(4\pi)^{3/2}}{V}(-1)^{m +m'}(-i)^{l_1 + l_2 +l_1' + l_2'} \nonumber\\
&\times \int r^2 dr\sum_{l_q l_p l_k} \frac{1} {\sqrt{(2l_q + 1)(2l_p + 1)(2l_k +1)}}\nonumber\\
&\times \sum_{J_1 J_2 J_3} \mathcal{D}_{J_1 J_2 J_3}\mathcal{C}_{J_1 J_2 J_3}\left(\begin{array}{ccc}
J_1 & J_2 & J_3\\
0 & 0 & 0
\end{array}\right)\nonumber\\
&\times \Bigg\{ \xi_{l_k}(r)\bigg[ w_1 f^{l_q}_{J_1 l_1 l_1'}(r; r_1, r_1') f^{l_p}_{J_2 l_2 l_2'}(r; r_2, r_2') \nonumber\\
&+ w_2 f^{l_q}_{J_1 l_1 l_2'}(r; r_1, r_2') f^{l_p}_{J_2 l_2 l_1'}(r; r_2, r_1')\bigg] + 
\left(\begin{array}{ccc}
J_1 & J_2 & J_3\\
S_1 & S_2 & S_3
\end{array}\right)\nonumber\\
&\times \bigg\{ f^{l_q}_{J_1 l_1}(r; r_1) \bigg[w_3 f^{l_p}_{J_2 l_2 l_2'}(r; r_2, r_2') f^{l_k}_{J_3 l_1'}(r; r_1')\delta^K_{S_1 -m, S_3 -m'} \nonumber\\
&+ w_4 f^{l_p}_{J_2 l_2 l_1'}(r; r_2, r_1') f^{l_k}_{J_3 l_2'}(r; r_2')\delta^K_{S_1 -m, S_3 m'} \bigg]\nonumber\\
&+  f^{l_p}_{J_2 l_2}(r; r_2) \bigg[w_5 f^{l_q}_{J_1 l_1 l_2'}(r; r_1, r_2') f^{l_k}_{J_3 l_1'}(r; r_1')\delta^K_{S_2 m, S_3 -m'} \nonumber\\
&+ w_6 f^{l_q}_{J_1 l_1 l_1'}(r; r_1, r_1') f^{l_k}_{J_3 l_2'}(r; r_2')\delta^K_{S_2 m, S_3 m'} \bigg] \bigg\}\Bigg\}.
\label{eqn:covar_w_wts}
\end{align}
In the terms proportional to $w_3$ through $w_6$, we fix two of the spins using the Kronecker deltas, and the third is given in equation (\ref{eqn:wts}) to avoid showing arithmetic in the index of a Kronecker delta. We emphasize that equation (\ref{eqn:covar_w_wts}) is fully general because the power spectrum can always be expanded in a Legendre series.

We close by emphasizing that equation (\ref{eqn:covar_w_wts}), though it appears complicated, substantially reduces the computational burden of obtaining the covariance matrix. The weights are simply an enumerable set of constant coefficients, and the main work is computing the $f-$tensors. However, these are simply 2-D and 3-D integral transforms of the power spectrum, and once a set of them is computed it can be combined to form the full covariance as above. Furthermore, since $r_1, r_2$, $r_1'$ and $r_2'$ are binned, one can replace the spherical Bessel functions by their bin-averaged values, and the $f-$tensors need then only be obtained at the $\sim\!\Nbins^2$ combinations of bin centers rather than on a full grid in any of the $r_i$. A fine grid in $r$ is required since $r$ is subsequently integrated over. Thus in essence the $f-$ tensors are not even truly 2-D or 3-D transforms, but rather 1-D cross a small set of bin center combinations. Equation (\ref{eqn:covar_w_wts}) therefore represents the reduction of the 12-D integral naively required for the covariance to a sum over a number of roughly 1-D integrals.

\subsection{Covariance with the Kaiser formula power spectrum}
\label{subsec:kaiser_covar}
We now observe that in the Kaiser approximation, where there is a single line of sight to the entire survey (\citealt{Kaiser1987}), and using linear perturbation theory, the redshift-space density rescales the real-space density as 
\begin{align}
\tilde{\delta}_s(\vec{k}) = (1+\beta \mu^2)^2\tilde{\delta}(\vec{k}),
\label{eqn:kaiser}
\end{align}
where $\beta = f/b_1$, with $f=d\ln D/d\ln a$ the logarithmic derivative of the linear growth rate $D$ with respect to scale factor $a$ and $b_1$ the linear bias. $\mu = \hat{k}\cdot \hat{n}$, where $\hat{n}$ is the line of sight to the survey. In this approximation, the $k$-dependence of all three moments of the anisotropic 2PCF is the same: it is just the power spectrum. The multipole moments simply have different constant prefactors, i.e.
\begin{align}
&c_0 = 1+\frac{2\beta}{3} + \frac{\beta^2}{5}\nonumber\\
&c_2 = \frac{4\beta}{3}+\frac{4\beta^2}{7}\nonumber\\
&c_4 = \frac{8\beta^2}{35},
\label{eqn:power_mult_wts}
\end{align}
where $c_0$ is for the monopole, $c_2$ for the quadrupole, and $c_4$ for the hexadecapole. 

This approximation permits considerable simplification of the covariance equation (\ref{eqn:covar_w_wts}). The upper index may be eliminated from all $f$-tensors, $\xi_{l_k}$ becomes a transform of the power spectrum against $j_{l_k}(kr)$ with a prefactor of $c_{l_k}$, and so a factor of $c_{l_q}c_{l_p}c_{l_k}$ is inserted within the sum over $l_q,l_p,$ and $l_k$. This step substantially reduces the number of $f$-tensors necessary to evaluate.

\subsection{Incorporating shot noise}
For any discrete sampling of the underlying continuous density field, a shot noise term inversely proportional to the survey number density $n$ will enter the covariance. As discussed in more detail in \cite{SE_3pt_alg}, this can be incorporated in the $f$-tensor framework simply by mapping $P(k)\to P(k)+1/n$, where in this case the shot noise term enters only the monopole moment of the power spectrum with respect to the line of sight. The shot noise can only enter the monopole because a discrete sampling effect should have no direction dependence. When the combinations $P+1/n$ are multiplied out, some of the terms will involve only $1/n$. As also detailed in \cite{SE_3pt_alg}, the $f$-tensors proportional to these terms can then be evaluated in closed form; this might be used to accelerate covariance matrix evaluations if desired.

\subsection{Reduction to the isotropic covariance}
As a check, we reduce the result equation (\ref{eqn:covar_w_wts}) to the isotropic 3PCF covariance by setting $l_1 = l_2 = l$, $l_1' = l_2' = l'$, and $l_q = l_p = l_k =0$, as well as summing over $m$ and $m'$. In this limit, we find
\begin{align}
&\mathcal{H}^{0 S_1 m m'}_{0 J_1 l l'} = \frac{1}{4\pi}\sqrt{(2J_1 + 1)(2l+1)(2l'+1)}\nonumber\\
&\times \left(\begin{array}{ccc}
J_1 & l & l'\\
0 & 0 & 0
\end{array}\right)
\left(\begin{array}{ccc}
J_1 & l & l'\\
-S_1 & -m & -m'
\end{array}\right)
\end{align}
where we used \citet{NIST_DLMF} 34.3.1 to evaluate two 3j-symbols that had zeros in one column (the full form of $\mathcal{H}$ is in \nameref{sec:appendix_A}).  Inserting this result (and its analog of the same form for different spins) into $w_1$, employing 3j-symbol identities \citet{NIST_DLMF} 34.3.10 and 34.3.18, and summing over $m$ and $m'$, we find 
\begin{align}
w_1 =\left(\begin{array}{ccc}
J_1 & l & l'\\
0 & 0 & 0
\end{array}\right)^2.
\end{align}
Applying the same manipulations, we find $w_2 = w_1$. 

The other weights are slightly more complicated. Thus we first present them prior to summing over $m$ and $m'$. For $w_3$, we find
\begin{align}
&w_3 = \frac{(-1)^{-m'-m}}{(4\pi)^2}\sqrt{(2J_2+1)(2l+1)(2l'+1)}\nonumber\\
&\times \left(\begin{array}{ccc}
J_2 & l & l'\\
0 & 0 & 0
\end{array}\right)
\left(\begin{array}{ccc}
J_2 & l & l'\\
-S_2 & m & m'
\end{array}\right);
\end{align}
$w_4$ is the same save for switching $m' \to -m'$ in the above. For $w_5$ we find
\begin{align}
&w_5 = \frac{(-1)^{m' + m}}{(4\pi)^2}\sqrt{(2J_1+1)(2l+1)(2l'+1)}\nonumber\\
&\times \left(\begin{array}{ccc}
J_1 & l & l'\\
0 & 0 & 0
\end{array}\right)
\left(\begin{array}{ccc}
J_1 & l & l'\\
-S_1 & -m & m'
\end{array}\right);
\end{align}
$w_6$ is the same save for switching $m' \to -m'$ in the above. 

Inserting $w_3$ through $w_6$ into equation (\ref{eqn:covar_w_wts}), we note that the pre-factors of $(-1)^{\pm m \pm m'}$ in the weights cancel with the overall pre-factor $(-1)^{m+m'}$. We now sum over $m$ and $m'$ with the appropriate replacements for $S_1, S_2$, and $S_3$ in the 3j-symbol that is a pre-factor of these terms. We then invoke orthogonality of the 3j-symbols summed over spins (\citealt{NIST_DLMF} 34.3.18) to find that the sums of products of spin-dependent symbols yield unity. We identify $J_1$ or $J_2$ as appropriate with $l_2$ of \cite{SE_3pt_alg} equation (65) for the isotropic covariance (they are simply dummy momenta coupling $l$ and $l'$). Finally, the pre-factor $(-i)^{l_1+l_2+l_1'+l_2'}$ in equation (\ref{eqn:covar_w_wts}) becomes $(-1)^{l+l'}$, and we multiply our covariance by $(4\pi)^2/[(2l+1)(2l'+1)]$, incorporating the pre-factor in equation (\ref{eqn:full_3pcf_est}) for each of the two 3PCFs forming the covariance. Simplifying what results reduces our anisotropic covariance to the isotropic covariance of \cite{SE_3pt_alg}.

\subsection{Symmetrization}
\label{subsec:covar_symm}
We now discuss how to cast the covariance matrix of the 3PCF's symmetrized harmonic coefficients in terms of the covariance of the unsymmetrized coefficients. 

We desire $\left<\bar{\zeta}_{l_1 l_2}^{m} \bar{\zeta}_{l_1' l_2'}^{m'} \right>$ where here we take it that $m\geq 0$ and $m'\geq 0$. Rewriting the symmetrized coefficients in terms of the unsymmetrized ones using equation (\ref{eqn:zeta_symm_defn}) and multiplying out, we obtain
\begin{align}
\left<\bar{\zeta}_{l_1 l_2}^{m} \bar{\zeta}_{l_1' l_2'}^{m'} \right>=&
\left<\zeta_{l_1 l_2}^{m}\zeta_{l_1'l_2'}^{m'} \right> + \left(1-\delta^{\rm K}_{m'0}\right)\left<\zeta_{l_1 l_2}^{m}\zeta_{l_1'l_2'}^{-m'} \right> \nonumber\\
&+ 
\left(1-\delta^{\rm K}_{m0}\right)\left<\zeta_{l_1 l_2}^{-m}\zeta_{l_1'l_2'}^{m'} \right>\nonumber\\
&+\left(1-\delta^{\rm K}_{m0}\right)\left(1-\delta^{\rm K}_{m'0}\right)\left<\zeta_{l_1 l_2}^{-m}\zeta_{l_1'l_2'}^{-m'}\right>.
\label{eqn:mult_out}
\end{align}
We notice that the last term is the complex conjugate of the first term for $m$ and $m'>0$, and that for $m$ and $m'=0$ it drops out (in this case the first term is its own complex conjugate).  Similarly, the third term is the complex conjugate of the second term for non-zero spins. This confirms our expectation that the covariance be real given that the symmetrized 3PCF coefficients are real. These observations mean that the complex conjugate pairs can be added to reduce the symmetrized coefficients' covariance to
\begin{align}
\left<\bar{\zeta}_{l_1 l_2}^{m} \bar{\zeta}_{l_1' l_2'}^{m'} \right>=&
\left[2-\delta^{\rm K}_{m'0}- \delta^{\rm K}_{m0}+\delta^{\rm K}_{m0}\delta^{\rm K}_{m'0} \right]\;{\rm Re}\;\left<\zeta_{l_1 l_2}^{m} \zeta_{l_1' l_2'}^{m'} \right>\nonumber\\
&+\left[2-\delta^{\rm K}_{m'0}- \delta^{\rm K}_{m 0}\right]\;{\rm Re}\;\left<\zeta_{l_1 l_2}^{m} \zeta_{l_1' l_2'}^{-m'}\right>,
\end{align}
where to add the second and third terms of equation (\ref{eqn:mult_out}) we used that 
\begin{align}
{\rm Re}\;\left<\zeta_{l_1 l_2}^{m} \zeta_{l_1' l_2'}^{-m'}\right> = {\rm Re}\;\left<\zeta_{l_1 l_2}^{-m} \zeta_{l_1' l_2'}^{m'}\right>
\end{align}
because the product within the expectation value on the lefthand side above is the complex conjugate of that on the righthand side above.

\section{Conclusions}
\label{sec:conclusions}

We have presented an algorithm for tracking the full 5-D anisotropic 3PCF. Assuming the RSD have azimuthal symmetry about the line of sight, the 3PCF depends on the three ``internal'' triangle parameters $r_1$, $r_2$, and $\hat{r}_1\cdot\hat{r}_2$ and the angles of two of the triangle sides to the line of sight. Here, we have traded these parameters for an equivalent 5-D representation---two side lengths $r_1$ and $r_2$, two total angular momenta $l$ and $l'$, and a spin $m$---and constructed the mixed spherical harmonic coefficients of the anisotropic 3PCF to capture its angle- and orientation-dependence. Our reformulation of the problem renders the density field integrals for these coefficients factorizable, fundamentally reducing the scaling of the problem from an $\mathcal{O}(N^3)$ triplet count to an $\mathcal{O}(N^2)$ pair count, with $N$ the number of galaxies in the survey. 

In addition to its speed, the algorithm presented here has three other significant advantages. First, it allows use of a rotating line of sight for the 3PCF, more accurate than assuming a single line of sight to the entire survey. The line of sight to the galaxy triplet is taken to be the vector to one of the three triplet members; this is the analog of the Yamamoto estimator for the anisotropic 2PCF or power spectrum. Furthermore, we have shown how these rotations can be done after computation of the spherical harmonic moments, enabling use of FFTs to evaluate the anisotropic 3PCF. Second, the spherical harmonic basis permits straightforward edge correction, an essential step to remove spurious signal generated by the survey geometry rather than the underlying galaxy clustering. Third, the basis enables computation of the covariance matrix under the assumption of a GRF density described by a power spectrum with multipole moments with respect to the line of sight. In the Kaiser approximation of a flat-sky, single line of sight and linear perturbation theory, the power spectrum's multipole moments become particularly simple, further accelerating evaluation of the covariance.

We note that the basis of spherical harmonic moments advocated here is a compression of the full redshift-space anisotropic 3PCF, as formally an infinite number of $\ell$ and $\ell'$ are required to model an arbitrary function of two directions. However, as shown in \cite{SE_RV_sig} and \cite{RSD_model}, in practice for the isotropic 3PCF, a finite, small number of multipoles contains the bulk of the information, at least on scales sufficiently large to be well-modeled by perturbation theory. Given the structure of the anisotropic bispectrum in perturbation theory (e.g. \citealt{Rampf2012}), we expect this conclusion will hold for the anisotropic 3PCF as well.  Thus we believe that the spherical harmonic basis is a parsimonious yet effective compression of the full anisotropic 3PCF.

There has been some work on modeling the anisotropic component of the bispectrum: \cite{Scoccimarro1999} computes the tree-level redshift space bispectrum via Eulerian Standard Perturbation Theory (SPT). Notably, the tree-level Eulerian and Lagrangian predictions for the redshift-space bispectrum (and hence 3PCF, since it is just the inverse Fourier Transform) agree (\citealt{Rampf2012}). This latter work presents the full, unaveraged tree-level bispectrum prediction. In contrast, \citet{Scoccimarro1999} modeled an average of the 5-D redshift-space bispectrum over rotations of one wavevector about the other, reducing it to a 4-D function.  That work began with the basis of spherical harmonics for the bispectrum's orientation dependence times coefficients depending on the three wavevector magnitudes. The averaging then reduced this to a basis of Legendre polynomials in the angle between the unaveraged side and the line of sight times the same wavevector-magnitude-dependent coefficients. 

More recently, \cite{Gagrani2017} compared the information content of the full redshift-space bispectrum to that of its 4-D reduction computed in this way, showing that at the level of the Fisher matrix most of the information is retained after this averaging.  The basis we propose in this work could easily be used for the bispectrum as well, but it is sufficiently different from the basis used in \citet{Scoccimarro1999} and \citet{Gagrani2017} that it is not clear how one would average over rotations about one triangle side if so desired.  

We note that two important papers in the development of spherical harmonics and Legendre polynomials for the bispectrum and 3PCF are \cite{Verde2000} and \cite{Szapudi2004}. The former proposed expansion of the projected galaxy density field in spherical harmonics, showed  that the isotropic projected bispectrum can be written as a Legendre series, and noted that this formalism also covers the full 3-D case. However it did not discuss the anisotropic 3PCF or bispectrum. The latter proposed expanding the full 3-D 3PCF or bispectrum in Legendre polynomials, but again did not discuss anisotropy. 

Here we have presented the mathematical formalism for our anisotropic 3PCF algorithm; in a companion papers, we discuss in detail an implementation suitable for massive-scale high-performance computing (\citealt{Friesen17}). In particular, we modified a codebase originally developed in \cite{SE_3pt_alg} to track the anisotropic clustering as discussed here. We optimized this code to run on the Cray XC40 system Cori at Lawrence Berkeley National Laboratory's National Energy Research Supercomputing Center (NERSC), which comprises roughly 10k nodes, each with 68 compute cores. Running it on the largest available galaxy simulation, Outer Rim (\citealt{Habib2016}), with 2 billion haloes, we computed the anisotropic 3PCF out to $200\;\Mpch$ in 1070 seconds on 9636 nodes, achieving $5.06\;{\rm PFLOPS}$ sustained. At peak, the code achieved $9.8\;{\rm PFLOPS}$, roughly $39\%$ of peak performance, but $80\%$ given the instruction mix the algorithm requires.\footnote{For this implementation, we pre-rotated the galaxies, which while computationally less efficient is simpler to code than the post-rotation approach also presented here. We expect incorporating this latter optimization will only improve runtimes, though we do not expect significant gains because the pre-rotation is a small part of the total work in the current code, as discussed in \S\ref{subsec:rotating}.} This speed means that any anisotropic 3PCF computation for galaxy surveys of sizes available in the next decade is practical even on current computing resources. Indeed, the anisotropic 3PCF out to $R_{\rm max}=200\;{\rm Mpc}$ for all galaxies in the observable Universe ($\sim$$100$ billion) is computable in a few days with the algorithm on Cori. The speed is further important because a full 3PCF analysis requires computing many random catalogs' 3PCF for edge correction, and many mocks' 3PCF for pipeline testing, model and covariance matrix verification, and fitting the free parameters (volume and shot noise) of the covariance matrix. 

Future work will be translating the predictions of \cite{Rampf2012} into the spherical harmonic basis for direct comparison with the output of the algorithm. What is clear by inspection is that the isotropic part of the 3PCF can only generate $l=l'$ couplings, so any ``off-diagonal'' couplings $l\neq l'$ isolate $\mathcal{O}(f)$ contributions. Thus in principle these couplings provide a robust window on the growth rate. However, in practice, anisotropies in the survey mask can couple isotropic coefficients of the measured 3PCF to anisotropic coefficients of the edge-corrected 3PCF, and vice versa. More detailed analysis with mock catalogs and a realistic survey geometry will thus need to be conducted to fully quantify this coupling.

Nonetheless, we believe the algorithm presented here will enable precise, robust measurement of the growth rate of structure with the anisotropic 3PCF much as is already done with the anisotropic 2PCF and power spectrum.  Thus far, the anisotropic 3PCF has not been measured. Therefore the next step for future work is applying this algorithm to data. Numerous suitable samples exist already, such as the SDSS DR12 BOSS CMASS sample used in \cite{SE_Full3PCF_BAO} and \cite{Gil_Marin2017}) or, if smaller volume but higher number density were desired, VIPERS (\citealt{vipers_dr}). Ongoing and future surveys such as eBOSS \citep{Dawson2016} and DESI \citep{Levi13} will provide even larger, richer datasets to which to apply this algorithm.

Measuring the growth rate of structure via RSD is both an important lever on the cosmological parameters and a key test of our theory of gravity. While the anisotropic 2PCF and power spectrum already will probe it to extremely high precision with next-generation surveys, any additional sources of information can only strengthen our understanding of these two fundamental areas.  

Further, breaking the degeneracy between $\sigma_8$, $f$ and galaxy biasing is a challenging problem that requires several different observables to fully address. In addition to its importance for cosmology, extracting precise bias measurements will shed new light on galaxy formation. 

Importantly, measuring the anisotropic 3PCF requires no additional data over what is already used for the anisotropic 2PCF; the challenge is purely computational, not observational. Thus the algorithm presented here should enhance the scientific value of redshift survey data per dollar spent on telescope time.

\section*{Acknowledgments}
ZS especially thanks Shirley Ho for encouragement on the computing side of this study and Anya Nugent and Martin White for careful reads of the manuscript. We also thank Douglas Finkbeiner, Stephen Portillo, Natalie Roe, Roman Scoccimarro, Uro\v s Seljak, and David Weinberg for useful discussions. Support for this work was provided by the National Aeronautics and Space Administration through Einstein Postdoctoral Fellowship Award Number PF7-180167 issued by the Chandra X-ray Observatory Center, which is operated by the Smithsonian Astrophysical Observatory for and on behalf of the National Aeronautics Space Administration under contract NAS8-03060. ZS also acknowledges support from a Chamberlain Fellowship at Lawrence Berkeley National Laboratory (held previously to the Einstein) and from the Berkeley Center for Cosmological Physics. DJE acknowledges support as a Simons Foundation Investigator and
from U.S. Department of Energy grant DE-SC0013718.




\bibliographystyle{mnras}
\bibliography{} 




\section*{Appendix A}
\label{sec:appendix_A}
In this Appendix, we collect some important definitions and properties
of Legendre polynomials and spherical harmonics. A product of two Legendre polynomials can be linearized into a sum as \citep{Adams78}
\begin{align}
\mathcal{L}_k(\mu)\mathcal{L}_{k'}(\mu) = \sum_J (2J+1) \left(\begin{array}{ccc}
k & k' & J\\
0 & 0 & 0
\end{array}\right)^2\mathcal{L}_J(\mu),
\end{align}
where the sum's range is set by $|k-k'|\leq J\leq k+k'$ (the triangularity condition on the 3j-symbol).

The integral of a Legendre polynomial can be obtained using the recursion
\begin{align}
\mathcal{L}_n(\mu) = \frac{1}{2n+1}\frac{d}{d\mu}\left[\mathcal{L}_{n+1}(\mu) - \mathcal{L}_{n-1}(\mu) \right]
\end{align}
to rewrite $\mathcal{L}_n(\mu)$ as an exact differential.

The Gaunt integral is defined
\begin{align}
&\mathcal{G}_{l_1 l_2 l_3}^{m_1 m_2 m_3} \equiv \int d\Omega \;Y_{l_1 m_1}(\hat{r}) Y_{l_2 m_2}(\hat{r}) Y_{l_3 m_3}(\hat{r})  \nonumber\\
&=\mathcal{C}_{l_1 l_2 l_3} \left(\begin{array}{ccc}
l_1 & l_2 & l_3\\
0 & 0 & 0
\end{array}\right)
\left(\begin{array}{ccc}
l_1 & l_2 & l_3\\
m_1 & m_2 & m_3
\end{array}\right)
\end{align}
where $\mathcal{C}_{l_1 l_2 l_3}$ is defined in equation (\ref{eqn:delta_fn_defs}).

The integral of four spherical harmonics can be obtained by first linearizing two spherical harmonics into a sum over single spherical harmonics using the Gaunt integral. We begin with 
\begin{align}
Y_{l_1 m_1}(\hat{r}) Y_{l_2 m_2}(\hat{r}) = \sum_{LM} c_{LM}(l_1, l_2;m_1, m_2) Y_{LM}(\hat{r}),
\label{eqn:ylm_series}
\end{align}
where the coefficients $c_{LM}(l_1, l_2;m_1, m_2)$ are given by integrating both sides against $Y^*_{LM}(\hat{r})$ and invoking orthogonality, so that
\begin{align}
c_{LM}(l_1, l_2;m_1, m_2) = (-1)^M\mathcal{G}_{l_1 l_2 L}^{m_1 m_2 -M}.
\label{eqn:cLM}
\end{align}
We then have
\begin{align}
&\mathcal{H}_{l_1 l_2 l_3 l_4}^{m_1 m_2 m_3 m_4} \equiv \int d\Omega\; Y_{l_1 m1}(\hat{r}) Y_{l_2 m_2}(\hat{r}) Y_{l_3 m_3}(\hat{r}) Y_{l_4 m_4}(\hat{r})\nonumber\\
&= \sum_{L} (-1)^M\mathcal{G}_{l_1 l_2 L}^{m_1 m_2 -M} \mathcal{G}_{L l_3 l_4}^{M m_3 m_4}
\end{align}
by inserting equations (\ref{eqn:ylm_series}) and (\ref{eqn:cLM}) into the first line above and then integrating. We note that there is no sum over $M$ because it is set by the zero-sum rule on the spins enforced by the Gaunt integrals: $m_1 + m_2 = M$. We further note that the sum over $L$ has compact support because of the triangle rules on total angular momenta: $|l_1 -l_2|\leq L\leq l_1 + l_2$ and the same constraint holds replacing  $l_1\to l_3$ and $l_2\to l_4$.

\section*{Appendix B}
\label{sec:appendix_B}
Having focused this paper on the $\zeta^m_{l l'}$ parameterization, 
we here explain why an alternative, seemingly attractive parameterization
of the anisotropic 3PCF does not work. Specifically, by analogy with the anisotropic 2PCF, one might expect that a triple Legendre series in the angle of each triangle side to the line of sight and the internal angle enclosed by the triangle would be the correct basis for the anisotropic 3PCF. However, this approach is not workable, as we show below.  

First, we write out the basis about a particular galaxy at $\vec{x}$; the full anisotropic 3PCF would then be the average over $\vec{x}$ of these coefficients. We have
\begin{align}
&\hat{\zeta}(r_1, r_2; \hat{r}_1\cdot\hat{r}_2; \hat{r}_1\cdot\hat{n}, \hat{r}_2\cdot\hat{n};\vec{x}) = \nonumber\\
&\sum_{l l_1 l_2}\zeta_{l l_1 l_2}(r_1, r_2;\vec{x})\mathcal{L}_l (\hat{r}_1\cdot\hat{r}_2)\mathcal{L}_{l_1} (\hat{r}_1\cdot\hat{n})\mathcal{L}_{l_2} (\hat{r}_2\cdot\hat{n}),
\label{eqn:bad_basis}
\end{align}
with $\hat{n}$ the line of sight.

There are two ways to see the flaw in this basis. First, consider a triangle with zero opening angle, so that $\hat{r}_1\cdot\hat{r}_2 = 1$.  Then specifying the orientation of $\hat{r}_1$ with respect to the line of sight fully specifies that of $\hat{r}_2$: the Legendre polynomials in $l_1$ and $l_2$ are no longer independent. This lack of independence means that they do not form a basis.

A more formal way to see this issue is by considering how we would obtain the coefficients $\zeta_{l l_1 l_2}(r_1, r_2; \vec{x})$. Placing a particular galaxy at $\vec{x}$ and measuring the coefficients around it, we would attempt to invoke orthogonality by integrating over $d\Omega_{r_1}d\Omega_{r_2}$. To do so we would need to separate the Legendre polynomials on the righthand side of equation (\ref{eqn:bad_basis}) as
\begin{align}
&\mathcal{L}_l (\hat{r}_1\cdot\hat{r}_2)\mathcal{L}_{l_1} (\hat{r}_1\cdot\hat{n})\mathcal{L}_{l_2} (\hat{r}_2\cdot\hat{n})=\frac{(4\pi)^2}{(2l+1)\sqrt{(2l_1+1)(2l_2+1)}}\nonumber\\
&\times\sum_{L_1 L_2}\mathcal{G}_{l l_1 L_1}^{000}\mathcal{G}_{l l_2 L_2}^{000} Y_{L_1 0}(\hat{r}_1) Y_{L_2 0}^*(\hat{r}_2).
\label{eqn:separation}
\end{align}
We used the spherical harmonic addition theorem to expand each Legendre polynomial, and then that $\hat{n} = \hat{z}$ to eliminate the sums over $m_1$ and $m_2$, since only the $m_1 = m_2 = 0$ modes contribute for a spherical harmonic evaluated along the $z$-axis. We then linearized the resulting products of two spherical harmonics in $\hat{r}_1$ and $\hat{r}_2$ into a sum over one spherical harmonic in each using equation (\ref{eqn:ylm_series}) from \nameref{sec:appendix_A}, and substituted the explicit form for the linearization coefficients as it is particularly simple since all spins are zero.

Inserting equation (\ref{eqn:separation}) into equation (\ref{eqn:bad_basis}) and integrating both sides against spherical harmonics over $d\Omega_{r_1}d\Omega_{r_2}$,  we see that
\begin{align}
&\int d\Omega_{r_1}d\Omega_{r_2} \;\hat{\zeta}(r_1, r_2; \vec{x})\;Y^*_{L_1 M_1}(\hat{r}_1) Y_{L_2 M_2}(\hat{r}_2) \nonumber\\
&\propto\sum_{l l_1 l_2 m}\mathcal{G}_{l l_1 L_1}^{000}\mathcal{G}_{l l_2 L_2}^{000} \zeta_{l l_1 l_2}(r_1, r_2; \vec{x}) .
\end{align}
We see that we have failed to extract the desired coefficients $\zeta_{l l_1 l_2}$, but have only succeeded in measuring some weighted sum of them. If the $l, l_1$, and $l_2$ support of the anisotropic 3PCF were finite, we could measure a large number of integrals as on the lefthand side above and solve for each term in the sum. However, it is known that the multipole expansion of the isotropic 3PCF is formally infinite (\cite{SE_RV_sig}; \cite{RSD_model}), so $l$ ranges from zero to infinity and this approach is not possible. We note that this point is not at odds with our claim in \S\ref{sec:conclusions} that the multipoles are a parsimonious basis for the 3PCF. While the multipole expansion is formally infinite, as shown in \citet{RSD_model}, Figure 8, the higher multipoles do not have much new information over the lower (their side-length dependence largely converges to look similar for $l\gtrsim 5$), so the information content remains compact.


\bsp	
\label{lastpage}
\end{document}